\documentstyle[psfig]{mn}

\newcommand{\fracd}[2]{\mbox{$\frac{\displaystyle{#1}}{\displaystyle{#2}}$}}

\newcommand{\mstar}{M_\star}
\newcommand{\msun}{M_\odot}
\newcommand{\rsun}{R_\odot}
\newcommand{\mearth}{m_\oplus}
\newcommand{\mjup}{m_{\rm J}}

\newcommand{\mpl}{m_{\rm p}}
\newcommand{\rpl}{r_{\rm p}}
\newcommand{\tpl}{t_{\rm p}}
\newcommand{\upl}{u_{\rm p}}
\newcommand{\thetap}{\theta_{\rm p}}

\newcommand{\rh}{S_{\rm L}}
\newcommand{\ml}{M_{\rm L}}

\newcommand{\rs}{R_{\rm S}}
\newcommand{\us}{u_{\rm S}}

\newcommand{\re}{R_{\rm E}}
\newcommand{\te}{t_{\rm E}}

\newcommand{\thetae}{\theta_{\rm E}}
\newcommand{\thetas}{\theta_{\rm S}}
\newcommand{\murel}{\mu_{\rm rel}}

\newcommand{\tdone}{t_{\rm done}}
\newcommand{\texp}{t_{\rm exp}}
\newcommand{\tslew}{t_{\rm slew}}
\newcommand{\tread}{t_{\rm read}}
\newcommand{\tmin}{t_{\rm min}}
\newcommand{\tmax}{t_{\rm max}}

\newcommand{\nmax}{N_{\rm max}}

\newcommand{\fsource}{f_{\rm S}}
\newcommand{\fstar}{f_\star}
\newcommand{\fvega}{f_0}
\newcommand{\fsky}{f_{\rm sky}}
\newcommand{\fblend}{f_{\rm B}}
\newcommand{\taustar}{\tau_\star}
\newcommand{\tausky}{\tau_{\rm sky}}
\newcommand{\taublend}{\tau_{\rm B}}

\newcommand{\musky}{\mu_{\rm sky}}

\newcommand{\magstar}{m_\star}

\newcommand{\aeff}{A_{\rm eff}}
\newcommand{\dl}{D_{\rm L}}
\newcommand{\ds}{D_{\rm S}}

\newcommand{\msq}{{\rm m}^2}
\newcommand{\cmcm}{{\rm cm}^2}

\newcommand{\dd}{{\rm d}}

\newcommand{\gtsim}{\stackrel{>}{_\sim}}
\newcommand{\ltsim}{\stackrel{<}{_\sim}}

\begin{document}

\title[ Optimising Microlens Planet Searches ]
{ A Metric and Optimisation Scheme for Microlens Planet Searches}

\author[Horne, Snodgrass]
{ Keith Horne$^{1}$, Colin Snodgrass$^{2}$, Yianni Tsapras$^{3,4}$
\\	$^{1}$SUPA Physics and Astronomy,
	University of St.Andrews, North Haugh, St.Andrews KY16 9SS,
	Scotland, UK.
\\	(kdh1@st-and.ac.uk).
\\	$^{2}$European Southern Observatory,
	Alonso de Cordova 3107, Casilla 19001,
	Vitacura, Santiago 19, Chile.
\\	(csnodgra@eso.org).
\\	$^{3}$Las Cumbres Observatory Global Telescope Network,
	6740B Cortona Dr, Suite 102, Goleta, CA, 93117, USA.
\\	$^{4}$Astrophysics Research Institute,
	Liverpool John Moores University, Twelve Quays House, 
	Egerton Wharf, Birkenhead, CH41~1LD, UK.
\\	(ytsapras@lcogt.net).
}

\date{Accepted . Received ; 
in original form }

\maketitle

\begin{abstract}

OGLE~III and MOA-II are discovering 600-1000 Galactic Bulge microlens
events each year.  This stretches the resources available for intensive
follow-up monitoring of the lightcurves in search of anomalies caused by
planets near the lens stars. We advocate optimizing microlens planet
searches by using an automatic prioritization algorithm based on the
planet detection zone area probed by each new data point. This
optimization scheme takes account of the telescope and detector
characteristics, observing overheads, sky conditions, and the time
available for observing on each night. The predicted brightness and
magnification of each microlens target is estimated by fitting to
available data points. The optimisation scheme then yields a decision on
which targets to observe and which to skip, and a recommended exposure
time for each target, designed to maximize the planet detection
capability of the observations. The optimal strategy maximizes detection
of planet anomalies, and must be coupled with rapid data reduction to
trigger continuous follow-up of anomalies that are thereby found. A web
interface makes the scheme available for use by human or robotic
observers at any telescope. We also outline a possible self-organising
scheme that may be suitable for coordination of microlens observations
by a heterogeneous telescope network.

\end{abstract} 

\begin{keywords}
gravitational lensing, planetary systems, methods: observational
\end{keywords}

\section{Introduction}
\label{sec:intro}


 
Gravitational microlensing reveals stars and planets that
magnify the light from a background source star \cite{mp91}.
The wide-field
OGLE~III\footnote{\verb+http://www.astrouw.edu.pl/$\sim$ogle+}
\cite{u03} and 
MOA~II\footnote{\verb+http://www.phys.canterbury.ac.nz/moa/+}
surveys of Galactic Bulge starfields
discover $\sim600-1000$ microlensing events each year.
During these events, a background source star brightens and fades, sometimes 
by many magnitudes,
in $t_E\sim30\left(M_\star/0.3~\msun\right)^{1/2}$ 
days as the intervening $M_\star\sim0.1-1~\msun$ lens star
crosses near the line of sight.
A planet near the lens star 
acts as a smaller lens,
smaller by a factor $\left(\mpl/M_\star\right)^{1/2}$.
When appropriately placed, the planet can produce
a brief but easily detectable flash or dip in the lightcurve.
Such planet anomalies last $t_p\sim3\left(\mpl/\mjup\right)^{1/2}$ 
days,
thus a few days for Jupiters or a few hours for Earths.
The probability that the planet is detectable
is $P_{\rm det}\sim0.2\left(\mpl/\mjup\right)^{1/2}$
for ``cool planets'' in the ``lensing zone'',
$a\sim0.5-2~\re\sim1-4\left(M_\star/0.3~\msun\right)^{1/2}$~AU
\cite{gl92}.

When a planet anomaly is well sampled by observations,
its duration, timing, and shape determine the
mass ratio, $q = \mpl/M_\starÂ$ and the orbit size $a$ relative to the 
Einstein ring radius $\re$.
Roughly speaking, the planet anomaly's duration $t_p$ 
sets the mass ratio, $q\sim \left(t_p/t_E\right)^2$, 
and its time $t$, relative to the event peak at $t_0$,
measures the projected planet-star separation
$a\sin{\theta}/\re \sim (t-t_0)/t_E$.
The lens star's distance $D$ and mass $\mstar$, when
constrained by the event timescale $t_E$, are initially
uncertain to factors $\sim3$.
Several methods using finite-source effects, parallax, and proper
motion can further constrain the lens geometry to
establish $\mpl$, $a$ and $M_\star$ with higher accuracy \cite{g09}.
For example, high-resolution imaging several years after the event
can detect the lens star flux, colour and proper motion \cite{bag07}.
 
The $m^{1/2}$ dependence of Einstein ring sizes makes
microlensing more sensitive to low-mass planets than other methods.
The microlens signature of an Earth-mass planet
is brief, a few hours, but can be strong enough for easy 
detection \cite{br96,d+07} provided one is observing the 
right star at the right time.  
With a detection probability
$P_{\rm det} \sim 0.2 \left(\mpl/\mjup\right)^{1/2}$
\cite{gl92}, a 
dedicated survey monitoring $\sim10^3$ events 
with $<1$ hour sampling could reveal $\sim10~\eta_\oplus$ cool Earths,
if each lens star has $\eta_\oplus$ of them.
While this level of monitoring has not yet been achieved,
significant constraints on the abundance of large cool planets 
were established \cite{g+02,t+03,s+04} even before the first
secure microlens planet detection; $\eta_{\rm Jup}<20$\%.

A two-stage strategy is currently employed for microlens planet searches.
The OGLE~III and MOA~II teams use their dedicated wide-angle survey 
telescopes to discover the microlens events.  Follow-up 
teams then deploy networks of small narrow-field telescopes distributed in 
longitude to obtain more intensive coverage of the most promising of those.
Two primary strategies are currently advocated and followed by the follow-up 
teams.
A strong focus on high magnification events, which have the highest 
probability of revealing planets, is advocated
\cite{gs98,r+02} and put into practice by $\mu$FUN
\footnote{\verb+http://www.astronomy.ohio-state.edu/$\sim$microfun/+}.
The high-magnification events are often identified a few days in advance, 
permitting the rapid mobilisation of many telescopes to cover the peak of 
the lightcurve as intensively as possible.
The PLANET\footnote{\verb+http://planet.iap.fr+}
Collaboration \cite{a+98} deploys a network of small ground-based 
telescopes to achieve quasi-continuous coverage of the most promising 
events.
This effort has been joined by 
RoboNet\footnote{\verb+http://robonet.lcogt.net/+}
\cite{b+07,t+09}, using three 2~m 
robotic telescopes.
A much larger robotic telescope network is being laid out by 
LCOGT\footnote{Las Cumbres Observatory Global Telescope.
\verb+http://lcogt.net+} in the
next few years \cite{t+09}.  With the prospect of 
this network of 24 0.4m and 18 1.0~m
robotic telescopes contributing to microlens planet searches,
automated strategies will be increasingly important to effectively 
organise the follow-up observations.

The OGLE-2002-BLG-055 lightcurve has one good data point that is 0.6~mag 
high. While this could be a planet anomaly \cite{jp02}, undersampling
prevents adequate characterisation of this event \cite{gh04}.
In the first secure characterisation of a microlens planet,
the lightcurve of OGLE-2003-BLG-235/MOA-2003-BLG-053
exhibits two fold caustics separated by 7 days, attributed to
lensing by a $\sim1.6~\mjup$ planet \cite{b+04}. 
The 2005 season revealed three microlens planets, 
$\sim3~\mjup$ OGLE-2005-BLG-071Lb \cite{u+05,d+09},
$\sim6~\mearth$ OGLE-2005-BLG-390Lb \cite{b+06},
and $\sim13~\mearth$ OGLE-2005-BLG-169Lb \cite{g+06}.
With two small planets among the first four microlens planet discoveries,
the abundance of small cool planets must be higher than that of the larger
cool Jupiters.
In 2006, the complex lightcurve of the high-magnification event 
OGLE-2006-BLG-109 revealed two planets with mass and orbital 
size ratios that are strikingly similar to
Jupiter and Saturn, scaled to a lower-mass ($\sim0.25~M_\odot$)
host star \cite{g+08}.
From the 2007 season, MOA-2007-BLG-192 \cite{b+08}
appears to be a brown dwarf with a $\sim3~\mearth$ planetary companion.
Other planets from 2007 are not yet published.
It appears reasonable on present evidence to expect increasing numbers of 
microlens planet discoveries, leading to detection of cool
Earth-mass planets within a few years, provided the capabilities
for intensive monitoring of OGLE~III and 
MOA~II events continues to improve.

This paper develops an optimal strategy
for reactive microlens planet searches that may help to increase
the planet discovery rate, particularly with dedicated 
telescope networks.
Section \ref{sec:lightcurves}
briefly reviews microlens lightcurves 
to define notation and establish a
few results for later use.
Section \ref{sec:zones}
employs numerical integrations and scaling laws
to quantify the detection zone area that we propose
as the metric of success for a microlens planet search.
Section \ref{sec:strategy}
develops the optimal observing strategy.
Section \ref{sec:discussion}
discusses several practical issues,
and outlines a possible self-organising scheme based on
continuously varying target priorities, that may be suitable
for coordinating microlens observations by a heterogeneous telescope 
network.
Section \ref{sec:summary} summarises the main results,
and describes our web interface to the PLOP (Planet Lens 
OPtimisation) algorithm.

\section{ Microlens Lightcurves }
\label{sec:lightcurves}

\subsection{ Point-Source Point-Lens (PSPL) Lightcurve }
\label{sec:pointlens}

During a microlensing event, light from a background 
source star reaches the Earth along paths that bend toward
an intervening lens star.
With perfect alignment of the observer, lens and source,
the observer sees the background star
as an Einstein ring of angular radius $\thetae$ centred on the lens.
A light ray with impact parameter $R$ bends 
toward the lens mass $M$ by a small angle
\begin{equation}
\alpha = \fracd{ 2\,\rh }{ R}
\ ,
\end{equation}
where $\rh=2\,G\,\ml/c^2$ is the Schwarzschild radius of the lens.
The point-mass gravitational lens has
strong spherical aberration, the effective focal length being
\begin{equation}    
f = \fracd{R}{\alpha} = \fracd{R^2}{2\,\rh}
\ .
\end{equation}
If $\dl$ and $\ds$ are the observer-lens and observer-source
distances, respectively, the lens formula of geometric optics is 
\begin{equation}
\fracd{1}{\dl} + \fracd{1}{\ds-\dl} = \fracd{1}{f} = \fracd{2\,\rh}{R^2}
\ .
\end{equation}
Solving for $R$ gives the radius of the Einstein Ring,
\begin{equation} 
\re = \left( 
	2\,\rh\ds\, X \left( 1-X \right)
	\right)^{1/2}
\ ,
\end{equation}
where $X\equiv \dl/\ds$ is the lens/source distance ratio, $0<X<1$.
The angular radius of the Einstein ring is
\begin{equation} 
\thetae = \fracd{\re}{\dl} =
\left( 2\,\rh\, \ds\, \fracd{1-X}{X} \right)^{1/2}
\ .
\end{equation}
For $\ml=\msun$, $\dl=5$~kpc, and $\ds=10$~kpc, 
the Einstein Ring radius $\re\approx4$~AU
corresponds to
$\thetae\sim0.8$~milli-arcseconds.

With imperfect alignment,
one ray on each side of the lens reaches the observer.
In this case, the lens equation is quadratic
with two distinct roots,
\begin{equation}
u_\pm = \fracd{u \pm \left( u^2 + 4 \right)^{1/2}}{2}
\ ,
\end{equation}
giving two image positions,
$u_+ = \theta_+/\thetae > 1 $ for
the major image and  $u_- = \theta_-/\thetae < 1 $ for the minor image,
in terms of $u=\theta/\thetae$ for the unlensed source.
Note for future reference that
\begin{equation}
\label{eqn:upmsq}
u_\pm^2 = \fracd{T \pm B}{2}
\ ,
\end{equation}
where $T = u_+^2 + u_-^2 = u^2+2$ and
$B = u_+^2 - u_-^2 = u \left( u^2+4 \right)^{1/2}$.

\begin{figure}
\begin{tabular}{c}
\psfig{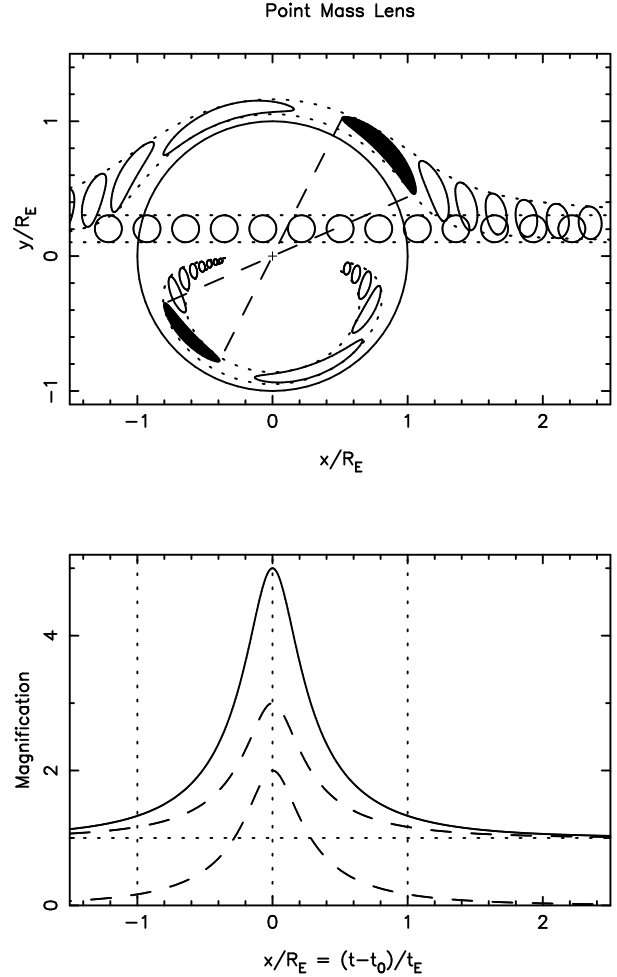}
\end{tabular}
\caption[] {\small
Top panel shows snapshots of the unlensed background star
and of the two distorted images of it that appear
on opposite sides of the lens during a micro-lensing event.
The major image passes over the top of the Einstein ring,
while the minor image executes a loop inside the Einstein ring.
The two images become compressed in radius but elongated in azimuth
as they approach the Einstein ring, resulting in a net magnification.
This produces the characteristic lensing lightcurves
shown in the bottom panel,
where dashed lines show the separate lightcurves of the two images,
$A_+(t)$ and $A_-(t)$,
and the solid line is their sum, $A(t)$.
Note that $A_+(t) = 1 + A_-(t)$.
\label{fig:pointlens}}
\end{figure}

Fig~\ref{fig:pointlens} shows the Einstein ring
and trajectories of the two images during a microlensing event.
On the lens plane perpendicular to the line of sight,
we define cartesian coordinates $x$ and $y$
with the origin at the lens star, 
the source star moving in the $+x$ direction
and crossing the $+y$ axis at closest approach.
In units of $\thetae$, the source-lens separation is
\begin{equation}
	u = \left( u_0^2 + u_x^2 \right)^{1/2}
\ ,
\end{equation}
with $u_0$ the separation at closest approach,
and
\begin{equation} 
u_x = \fracd{\mu\,(t-t_0)}{\thetae} = \fracd{ ( t - t_0 ) }{ \te }
\ ,
\end{equation}
where $\mu$ is the relative proper motion,
$t_0$ is the time of closest approach, 
and the event timescale,
\begin{equation}
\te = \fracd{\thetae}{\murel}
\ ,
\end{equation}
is the time to cross the radius of the Einstein ring.

\begin{figure}
\begin{tabular}{c}
\psfig{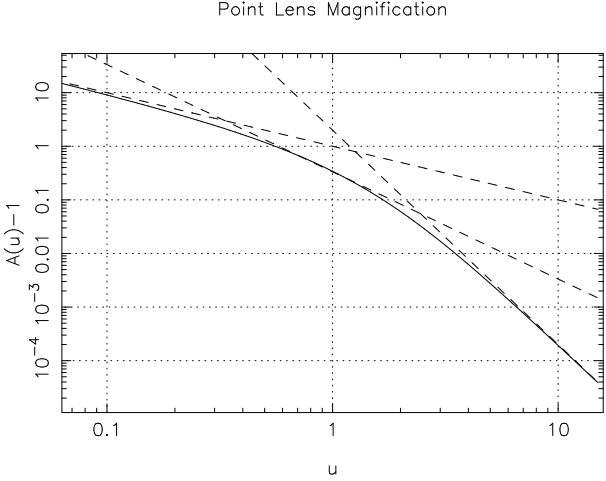}
\end{tabular}
\caption[] {\small
The point-source point-lens (PSPL) magnification
$A(u)$ compared with the power-law approximations
defined in Eqn~\ref{eqn:3pows}.
\label{fig:a(u)}}
\end{figure}

The image-lens separations satisfy
$u_+^2 u_-^2 = 1$, so that, 
as seen in Fig~\ref{fig:pointlens},
the major image at $u_+$ is always outside the Einstein ring,
while the minor image at $u_-$ remains inside.
The major image slides ``over the top'' of the Einstein ring,
while the minor image traces a loop inside the ring.
Both images become brighter as they approach the Einstein ring.
Each point on the disc of the source star maps
to a corresponding lensed position on the image.
The images are thus stretched in azimuth by a factor $u_\pm/u$
and squashed in radius by $\dd u_\pm/\dd u$.
With surface brightness conserved, the net magnification arising
from the increased solid angle is
\begin{equation}
\label{eqn:apm}
\begin{array}{rl}
	A_\pm & = 
\left| \fracd{u_\pm}{u} \right|
\left| \fracd{ \dd u_\pm }{ \dd u} \right| 
= \fracd{1}{2u} \fracd{\dd}{\dd u} \left[ u_\pm^2 \right]
\\ & = 
\fracd{u_\pm^2}{u_+^2 - u_-^2}
= \fracd{T \pm B}{2B} = \fracd{A \pm 1}{2}
\ .
\end{array}
\end{equation}
The image magnifications satisfy $A_+=1+A_-$,
and the total magnification is
\begin{equation}
\label{eqn:avsu}
A \equiv A_+ + A_- =
\fracd{u_+^2 + u_-^2}{u_+^2 - u_-^2} = \fracd{T}{B}
	= \fracd{ u^2 + 2 }{ u \left( u^2 + 4 \right)^{1/2} } 
\ .
\end{equation}
Since $u$ changes with time, this defines a
characteristic point-source point-lens (PSPL) lightcurve
(Fig~\ref{fig:pointlens}).
Power-law approximations (Fig.~\ref{fig:a(u)})
for large, intermediate, and small $u$ are
\begin{equation}
\label{eqn:3pows}
A(u) - 1 = \left\{
\begin{array}{cl}
	2 u^{-4}  & u \gtsim 2.5
\ ,
\\	u^{-2} / 3  & 0.3 \ltsim u \ltsim 2.5
\ ,
\\	u^{-1} & u \ltsim 0.3
\ .
\end{array}
\right.
\end{equation}
Note that $A(u)$ has the inverse:
\begin{equation} 
\label{eqn:ainv}
	u
	= \left( 2 \left[ \left( 1 - A^{-2}\right)^{-1/2} - 1 \right]
	\right)^{1/2}
\ .
\end{equation}

\subsection{Finite Source Size}
\label{sec:finitesource}

For a point source at high magnification,
as $u \rightarrow 0$, 
$A \rightarrow 1/u$ becomes formally infinite,
corresponding to the formation of an Einstein ring
of infinite magnification and infitesimal thickness.
This point-source approximation breaks down, however,
when the source star's finite size becomes important.
In Fig.\ref{fig:pointlens}, curvature of the highly magnified images
is already evident.
At still higher magnifications, the major and minor images
extend farther in azimuth, eventually touching each other
and merging to form an Einstein ring of finite width.
The magnification remains finite
due to the finite source size 
\cite{br96,d98}.
The source star's angular radius $\thetas=\rs/\ds$ 
becomes comparable to the angular radius $\thetae$ of the Einstein ring,
at
\begin{equation}
\begin{array}{rl}
\us & \equiv \fracd{\thetas}{\thetae}
	= \fracd{\rs}{\left( 2\,\rh\,\ds\,X\left(1-X\right) \right)^{1/2}}
\\ \\ &	= 0.0013 \left( \fracd{\rs}{\rsun} \right)
	\left( \fracd{M_L}{\msun} \right)^{-1/2}
	\left( \fracd{X}{1-X} \right)^{1/2}
\ .
\end{array}
\end{equation}
Finite-source effects set in at high magnification, $A\gtsim1/\us$,
thus (for $\ml\sim0.3\,\msun$) at $A\gtsim500$
for a main-sequence source star with $\rs \sim \rsun$, 
or already at $A\gtsim5$ for a giant source star with $\rs\sim100\,\rsun$.

Finite source effects are important for planet anomalies
when the source star's angular radius $\thetas$  exceeds
that of the planet's Einstein ring
\cite{br96}.
Since $\thetap = q^{1/2}\,\thetae$, the
finite source effect is important when
\begin{equation}
q \ltsim \left( \fracd{\thetas}{\thetae} \right)^2
\sim 2\times10^{-6}
	\left( \fracd{\rs}{\rsun} \right)^2
	\left( \fracd{M_L}{\msun} \right)^{-1}
	\left( \fracd{X}{1-X} \right)
\ .
\end{equation}
Since $\mearth = 3\times10^{-6}\msun$, this is
\begin{equation}
\mpl = q\,\ml \ltsim 0.6\, \mearth
	\left( \fracd{\rs}{\rsun} \right)^2
	\left( \fracd{X}{1-X} \right)
\ .
\end{equation}

For large source stars the anomaly from a small planet
can be smeared out and diluted in amplitude, rendering it undetectable.
On this basis, detection of Earth-mass planets is more favourable
with main sequence source stars \cite{br96}, though a detectable 
($\sim5$\%) signal can arise even when the source star is a giant 
\cite{d+07}, provided the planet's alignment with one of the image 
trajectories is favourable.
We do not consider finite-source effects further in this paper.

\subsection{ Binary Lens Anomalies }
\label{sec:binarylens}

A planet near the lens star acts like a small defect in the
gravitational lens.
If the planet is well away from the two image trajectories,
the light it deflects does not reach Earth.
In this case the planet has no measurable effect on the lightcurve
and thereby evades detection.
However, if the planet is close to one of the image trajectories,
its gravity can significantly perturb the bundle of light rays
that would otherwise reach Earth.
This distorts the image and changes the magnification to produce
a brief anomaly in the lightcurve.
The magnification curve $A_2(t)$
for a star+planet lens deviates by a factor $1 + \delta(t)$
from the corresponding
PSPL magnification curve $A_1(t)$:
\begin{equation}
	A_2(t)
	= A_1(t) \left( 1 + \delta(t) \right)
\ .
\end{equation}
This defines the planet anomaly $\delta(t)$,
which depends on three additional parameters:
the mass ratio $q$, and the coordinates, $x$ and $y$,
of the planet's projected position on the lens plane.

The planet anomaly may be brief but large.
The planet's Einstein ring radius is
\begin{equation} 
	\rpl \equiv \re\, q^{1/2}
\ .
\end{equation}
The planet anomaly may be large when one of the source
images passes closer to the planet than $\rpl$,
provided the source is not much larger than $\rpl$.
The duration of the planet anomaly is roughly the time it takes the
image to cross the diameter of the planet's Einstein ring,
\begin{equation} 
	\tpl \equiv \te\, q^{1/2}
\ .
\end{equation} 
Detecting the planet requires data points in the lightcurve
of sufficient accuracy and at the right time to
detect the anomaly produced as the image
passes by the planet.

\section{ Planet Detection Zones }
\label{sec:zones}

\subsection{ Definition of Detection Zone }
\label{sec:zone_def}

We define the ``detection zone'' as the region 
on the lens plane ($x$,$y$) 
where the lightcurve anomaly $\delta(t,x,y,q)$
is large enough to be detected or ruled out with high confidence
by the observations.
For $N$ data points with fractional accuracy $\sigma_i$
at times $t_i$, the detection zone is defined by
\begin{equation}
	\sum_{i=1}^N \left( \fracd{\delta(t_i,x,y,q)}{\sigma_i }\right)^2
> \Delta\chi^2
\ ,
\end{equation}
for some detection threshold $\Delta\chi^2$.
This detection threshold must be set high enough so that
noise affecting the observations does not produce
false triggers at an unacceptably high rate.
$\Delta\chi^2$ in the range 25 to 100 
corresponds to a $5\sigma$ to $10\sigma$
deflection in the lightcurve 
if the anomaly is confined to a single data point.

\begin{figure}
\begin{tabular}{c}
\psfig{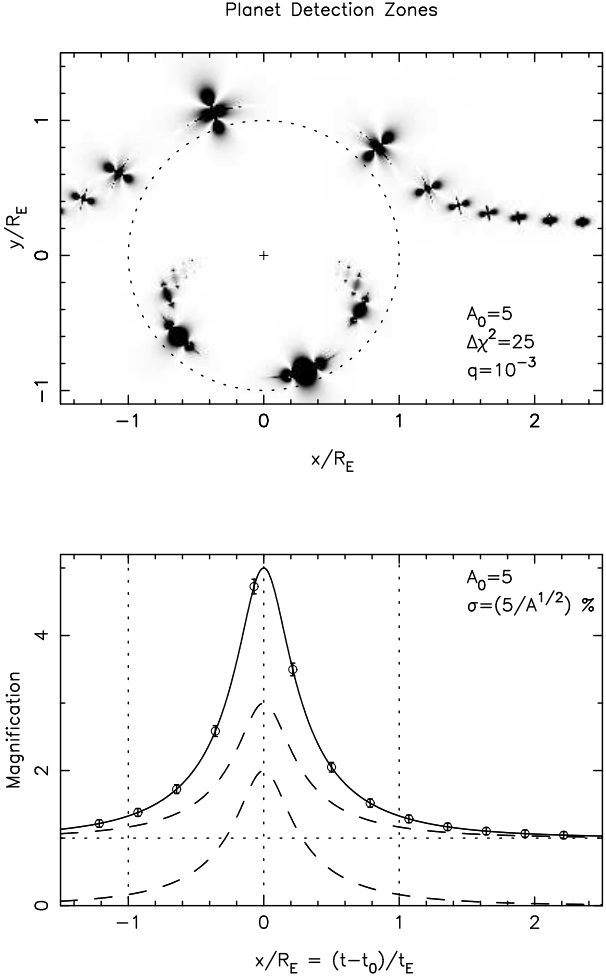}
\end{tabular}
\caption[] {\small
Detection zones on the lens plane indicate the
regions where a planet with mass ratio $q=m/M=10^{-3}$
is detected with $\Delta\chi^2>25$.
The lightcurve $A(t)$ has maximum magnification $A_0=5$,
and the accuracy of the measurements is $\sigma=(5/A^{1/2})$\%.
Each data point probes for planets close to
the two images of the background star.
The detection zone areas scale roughly as 
$\Omega \approx \re^2\,(2\,A-1)\,q\,/(\sigma\,\Delta\chi)$.
\label{fig:dzones}}
\end{figure}

Fig.~\ref{fig:dzones} 
highlights the detection zones 
for a planet with mass ratio $q=10^{-3}$
derived from a lightcurve $A(t)$ with
maximum magnification $A_0=5$.
Data points uniformly spaced in time sample the lightcurve
with an accuracy $\sigma=(5/A^{1/2})$\%, 
and the detection criterion is $\Delta\chi^2>25$.
Each data point probes for planets close to
the corresponding major and minor image positions.
If a planet is placed inside one of these detection
zones, the lightcurve at time $t_i$
is perturbed by $\delta>5\,\sigma=(25/A^{1/2})\%$.
Improving the accuracy of the data or 
increasing the mass of the planet
enlarges the size of the detection zone.

\subsection{ Numerical Evaluation of Detection Zone Areas }
\label{sec:zone_fine}

If the data points in the lightcurve are widely spaced,
as they are in Fig.~\ref{fig:dzones}, then
the detection zones arising from different data points
are well isolated from each other.
We may then evaluate numerically the area $\Omega$ of the
detection zone that is carved out by each data point.
This quantifies the planet discovery potential of
each data point.

\begin{figure*}
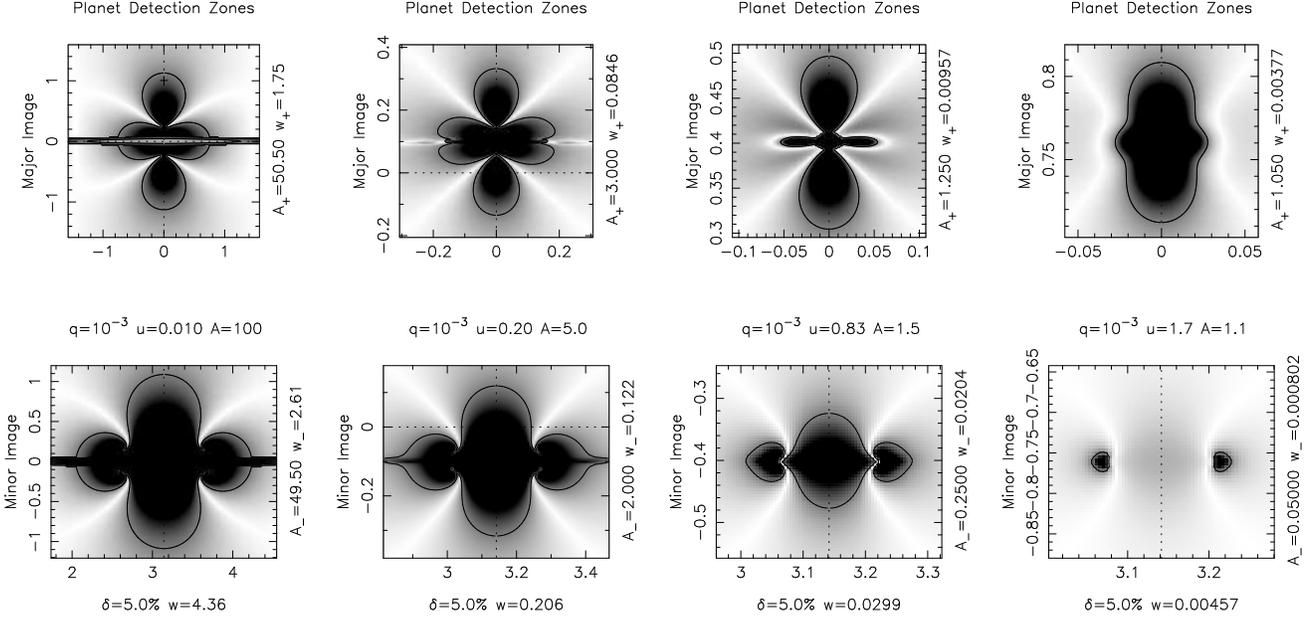

\begin{tabular}{cccc}
\psfig{file=horne_fig04a.ps,angle=0,width=4cm}
&
\psfig{file=horne_fig04b.ps,angle=0,width=4cm}
&
\psfig{file=horne_fig04c.ps,angle=0,width=4cm}
&
\psfig{file=horne_fig04d.ps,angle=0,width=4cm}
\end{tabular}
\caption[] {\small
Planet detection zones for mass ration $q=10^{-3}$
near the major and minor image positions
for data points at magnifications $A=100$, 5, 1.5, and 1.1.
The detection zones are symmetric around the image positions
in ($\theta$,$\ln{u})$ coordinates.
For $u\rightarrow\infty$ and $A\rightarrow1$
the major image detection zone becomes circular
and the minor image detection zone splits in two and vanishes.
\label{fig:zones}}
\end{figure*}

Fig.~\ref{fig:zones} shows a close-up of the detection zones
defined by this criterion.
The region displayed is chosen in advance from rough estimates
and is used for numerical evaluation of the dectection zone
area $\Omega$.
The cases shown illustrate how the detection zones
shrink and change shape as the magnification $A(u)$ declines
with increasing lens-source separation $u$.
The detection zone shapes are complicated.
At small $u$ and high $A$ they bear some resemblance to 
4-leafed clovers with radial and azimuthal lobes
straddling the image positions.
With increasing $u$, decreasing $A$, the azimuthal lobes
of the major image detection zone collapse radially.
The radial lobes then merge radially to form a circular
detection zone as $u\rightarrow\infty$.
On the minor image detection zone, the radial lobes merge and vanish,
leaving two isolated azimuthal lobes that shrink and vanish.

\subsection{ Scaling Laws for Detection Zone Areas }
\label{sec:scaling}

 It will be helpful to understand how planet detection zone
areas scale with the accuracy of the data, the
source magnification, and the mass of the planet.
If accurate scaling laws can be found, 
we may then avoid long numerical calculations to 
determine the detection zone area. In this section we develop
a useful analytic formula, and test it against
detailed numerical integrations.

Consider first a planet located quite far from the lens star,
affecting the major image at a time well before or
well after the stellar lensing event, 
when $u_+>>1$ and $A_+ \approx 1$.
In this case the planet and star act as independent lenses,
and a significant anomaly occurs when the major image
sweeps past the position of the planet.
If $z$ is the separation between the planet and the major image,
the planet magnifies the major image by a factor $A(\upl)$, 
where $\upl=z/\rpl$,
and $\rpl = q^{1/2}\,\re$ is the planet's Einstein ring radius.

\begin{figure}
\begin{tabular}{c}
\psfig{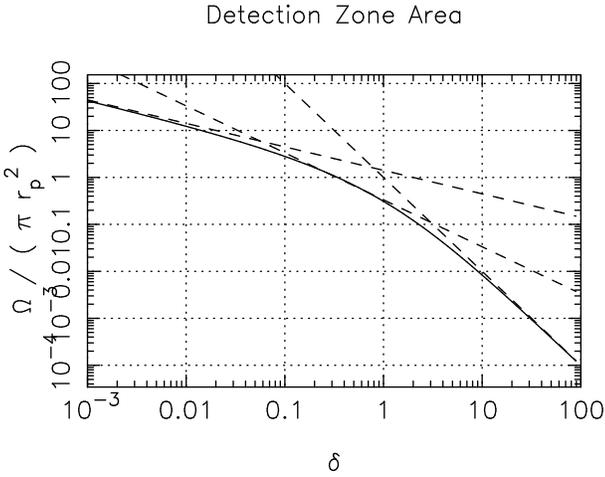}
\end{tabular}
\caption[] {\small
The detection zone area $\Omega$ 
scales with the area $\pi \rpl^2$
of the planet's Einstein ring,
and decreases with the size of the lightcurve anomaly 
$\delta$ that can be detected by the data.
The approximations (dashed lines) are those
defined in Eqn.~\ref{eqn:3omega}.
\label{fig:omega}}
\end{figure}

A data point with fractional uncertainty $\sigma$
can detect the anomaly $\delta$ when 
\begin{equation}
\label{eqn:delta}
	\delta = A(\upl)-1 > 
	\sigma
	\left( \Delta\chi^2 \right)^{1/2}
\ .
\end{equation}
This criterion corresponds to a circular detection zone around 
the major image at the time of the observation.
The radius of the detection zone is found by solving
Eqn.~(\ref{eqn:delta}) for $\upl$ and hence $z = \rpl \upl$.
Using Eqn.~(\ref{eqn:ainv}) to invert $A(\upl)$, 
the area of the detection zone is
\begin{equation}
\label{eqn:omega}
	\fracd{\Omega}{\pi \rpl^2} 
	= \left(\fracd{z}{\rpl}\right)^2
	= \left[
	2 \left( 1 - \left(1+\delta\right)^{-2} \right)^{-1/2} - 1
	\right]
\ .
\end{equation}
Using the approximations in Eqn.~(\ref{eqn:3pows}), the 
corresponding approximations for the detection zone area are
\begin{equation}
\label{eqn:3omega}
	\fracd{\Omega}{\pi \rpl^2} 
= \left\{
\begin{array}{cl}
	\left( 2 / \delta \right)^{1/2} & \delta \ltsim 0.05
\ ,
\\	\left( 3\ \delta \right)^{-1} & 0.05 \ltsim \delta \ltsim 3
\ ,
\\	\delta^{-2} & \delta \gtsim 3
\ .
\end{array}
\right.
\end{equation}
Fig.~\ref{fig:omega} indicates that the middle approximation 
predicts fairly accurately the detection zone area
for anomalies in the range $0.05 < \delta < 3$.
For $\Delta\chi^2=25$ this range corresponds
to fractional uncertainties $0.01 < \sigma < 0.6$,
quite appropriate for CCD data, giving
\begin{equation}
\fracd{\Omega}{R_E^2} = \fracd{\pi\,q}{3\,\delta}
\ .
\end{equation}

Now move the planet closer to the lens star.
The planet acts as a defect in the stellar lens, and
the resulting detection zone can have a quite complicated 
shape (Fig.~\ref{fig:zones}).
We guess that the detection zone area may scale roughly 
with the magnification $A$.
However, we find that $\Omega\propto2A-1$ is a better approximation:
\begin{equation}
	\fracd{\Omega}{\re^2} 
	 \approx \fracd{ \pi\,q}{ 3\,\delta} \left( 2\,A - 1 \right)
\ .
\end{equation}

Fig.~\ref{fig:zones} shows
that the detection zones are roughly symmetric
as a function of $\ln{u}$ and $\theta$ rather than $x$ and $y$.
It may therefore be more appropriate to express
detection zone areas using a $\dd\theta\,\dd\ln{u}$ metric,
evaluating
\begin{equation}
	w = \int 
	P( {\rm det} | u, \theta )\
	\fracd{\dd u\ \dd\theta}{u}
\ ,
\end{equation}
rather than
\begin{equation}
	\fracd{\Omega}{\re^2} = \int
	P( {\rm det} | u, \theta )\
	 u\,\dd u\,\dd\theta
	= \int 
	P( {\rm det} | x, y )\
	\dd x\, \dd y
\ .
\end{equation}

The $\dd\theta\,\dd\ln{u}$ metric
may be appropriate from a second perspective.
Exo-planet orbits should have random orientations,
so the planets should be uniformly distributed in $\theta$.
If their orbit size distribution is also roughly uniform in $\log{a}$,
then the planet distribution on the lens plane
will be roughly uniform in $\ln{u}$,
and the planet detection probability will decline
to zero long after the peak of an event,
rather than reaching a positive asymptotic value.
In effect the $\dd\theta\,\dd\ln{u}$ metric
recognizes a detection zone with area $\Omega$ 
as more likely to include a planet,
and therefore more valuable to us,
when it is measured at small $u$ and probes a larger
range of $\log{u}$.

\begin{figure}
\begin{tabular}{c}
\psfig{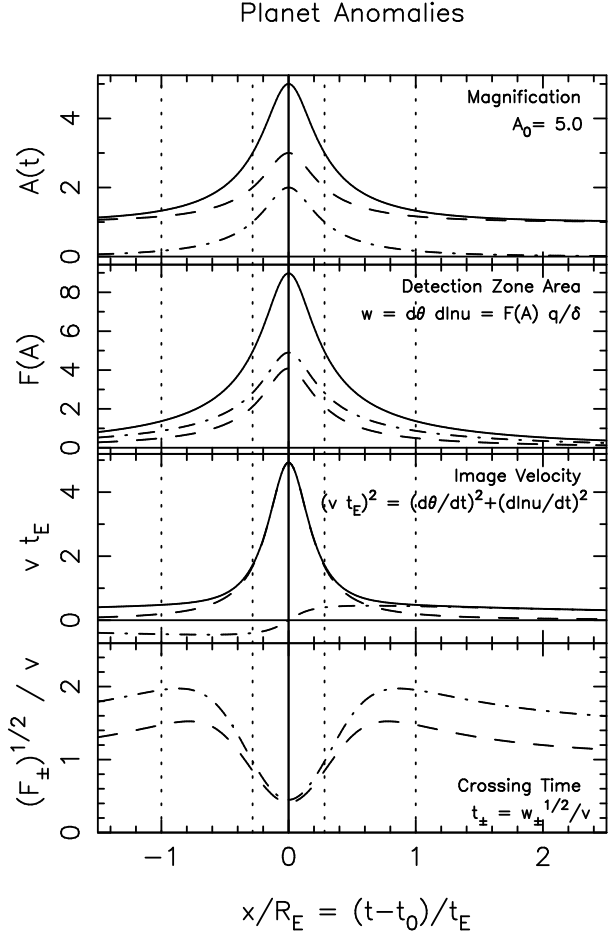}
\end{tabular}
\caption[] {\small
The magnification, detection zone area, image velocity, and
detection zone crossing time are shown for an event with peak magnification 
$A_0=5$.
The dashed and dash-dot curves are for the major and minor images 
respectively, except for the image velocity panel where the dashed 
curve is $\dd\theta/\dd t$ while the dash-dot curve is $\dd\ln{u}/\dd t$
for the major image.
\label{fig:afvt}}
\end{figure}

For small detection zones the two metrics are related by
\begin{equation}
	w \approx \fracd{\Omega}{\re^2\,u^2}
\approx \fracd{\pi\,q}{3\,\delta} 
\left( \fracd{2A-1}{u^2} \right)
\ .
\end{equation}
But we must be more careful to treat separately the major and minor image 
detection 
zones, surrounding the images at $u_+$ and $u_-$ respectively.
This gives
\begin{equation}
\label{eqn:wpm}
	w_\pm \approx \fracd{\pi\,q}{3\,\delta}\ F_\pm(A)
\ ,
\end{equation}
where for the major image at $u_+$
\begin{equation}
\label{eqn:fplus}
	F_+(A) \equiv \fracd{2\,A_+-1}{\left(u_+\right)^2}
	= \fracd{2\,T}{B\left(T+B\right)}
	= A\, \left( \fracd{A-1}{A+1} \right)^{1/2}
\ ,
\end{equation}
and for the minor image at $u_-$
\begin{equation}
\label{eqn:fminus}
	F_-(A) \equiv \fracd{2\,A_-}{\left(u_-\right)^2}
	= \fracd{2}{B}
	= \left(A+1\right)\, \left( \fracd{A-1}{A+1} \right)^{1/2}
\ .
\end{equation}
In deriving the above expressions, we used Eqns.~(\ref{eqn:upmsq}),
(\ref{eqn:apm}) and (\ref{eqn:avsu}) to write
\begin{equation}
\label{eqn:tb}
\begin{array}{rl}
	A_\pm = & \fracd{T \pm B}{2\,B}
\ , \hspace{3mm}
	\left(u_\pm\right)^2 = \fracd{ T \pm B }{2}
\ ,
\\ \\	T = & u_+^2 + u_-^2 = u^2 + 2
	= \fracd{2\,A}{ \left( A^2 - 1\right)^{1/2} }
\ , 
\\ \\	B = & u_+^2 - u_-^2 = u\, \left( u^2+4 \right)^{1/2}
	= \fracd{2}{ \left( A^2 - 1 \right)^{1/2} }
\ .
\end{array}
\end{equation}
The total detection zone area,
summing the detection areas of both images,
is
\begin{equation}
\label{eqn:w}
 w = \fracd{\pi\,q}{3\,\delta} F(A)
	= \fracd{\pi\,q\, F(A)}
	{ 3\, \left( \Delta\chi^2 \right)^{1/2}\, \sigma(\ln{A})}
\ ,
\end{equation}
with $\sigma(\ln{A})$ the fractional accuracy in measuring $A$,
$\Delta\chi^2$ the threshold for planet detection, and
\begin{equation}
\label{eqn:f}
\begin{array}{rl}
F(A) & \equiv F_+(A) + F_-(A)
\\ \\ & = \fracd{2}{B}\, \left( \fracd{2\,T+B}{T+B} \right)
	= \left( 2\,A+1 \right) \left( \fracd{A-1}{A+1} \right)^{1/2}
\ .
\end{array}
\end{equation}
The functions $F_\pm(A)$, for the separate images and $F(A)$ 
for the total detection zone area are plotted 
in Fig.~\ref{fig:afvt} for an event with $A_0=5$.

\subsection{Analytic vs Numerical Detection Zone Areas}

\begin{figure}
\begin{tabular}{c}
\psfig{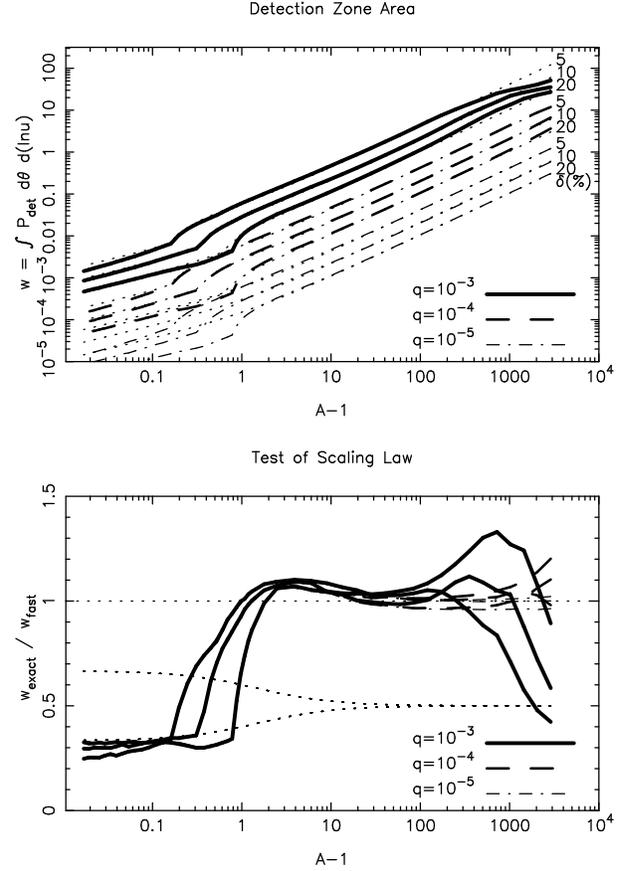}
\end{tabular}
\caption[] {\small
The detection zone areas $w$ in the $\dd\theta\,\dd\ln{u}$ metric
increase with magnification $A$,
increase with mass ratio $q$,
and decrease with the size of the lightcurve anomaly 
$\delta$ that can be detected by the data.
The fast analytic approximation (dotted)
defined in Eqn.~(\ref{eqn:w}) is compared with more exact numerical 
integrations.
Dotted curves in the lower panel give
fractional contributions of the major and minor images,
in the analytic approximation.
\label{fig:scaling}}
\end{figure}

Fig.~\ref{fig:scaling} compares numerically-integrated
detection zone areas with the analytic result in Eqn.~(\ref{eqn:w}).
The figure shows how $w$ depends on $A-1$
for three mass ratios, $q=10^{-3}$, $10^{-4}$ and $10^{-5}$,
and for three accuracies, 
$\delta=\sigma(\ln{A})\left(\Delta\chi^2\right)^{1/2}=5$, 10 and 20\%.

The analytic approximation clearly captures the main scaling,
$w\propto q\,A/\delta$, for $2<A<300$.
However, it is a rough guide rather than a superb approximation.
At moderately high magnifications, $10<A<300$,
the analytic and numerical results agree to within $\sim5$\%.
Here we have $F_+\approx\,F_-\approx(2\,A-1)/2$, 
the two images contributing roughly equally.
At intermediate magnifications, $2<A<10$, 
the analytic result is low by up to $\sim10$\%.
This $A$-dependent bias could be reduced by adjusting
the formula for $F(A)$ in Eqn.~(\ref{eqn:f}).
However, the accuracy is already sufficient for our
purposes in the range $2<A<300$. 
We discuss below the breakdown at larger and smller magnifications.

At very high magnifications, $A>300$, our analytic approximation breaks down 
because the  detection zones extend so far in azimuth that the major and minor 
image zones touch each other and merge together. 
Since the detection zones span a roughly equal range in $\theta$ and in $\ln{u}$,
the saturation in $\theta$ should set in when 
$w\sim\left(2\,\pi\,q\,A/3\,\delta\right)\gtsim\pi^2$, i.e.
\begin{equation}
	A \gtsim \fracd{3\,\pi}{2}\, \fracd{\delta}{q}
\ .
\end{equation}
We enter a new regime in which the azimuthally-merged detection zone may continue 
to expand in $\ln{u}$ but is saturated in $\theta$.
The slope should drop to $w \propto \left( q\,A/\delta \right)^{1/2}$, i.e.
\begin{equation}
	w\left( A \gtsim \fracd{\delta}{q} \right)
	\approx 2\,\pi \left( \fracd{2\,\pi\,q\,A}{3\,\delta} \right)^{1/2}
\ .
\end{equation}
These expectation are roughly consistent with the behaviour in 
Fig.~\ref{fig:scaling}.

Note that in the very-high magnification regime finite source effects will also
become important, altering the relationship between $u$ and $A$ with $u>\us$,
as discussed in Sec.~\ref{sec:finitesource}.
As we neglect both effects in our analysis, our scaling law applies
only up to a maximum magnification $A\ltsim\delta/q\sim300$.

\begin{figure}
\begin{tabular}{c}
\psfig{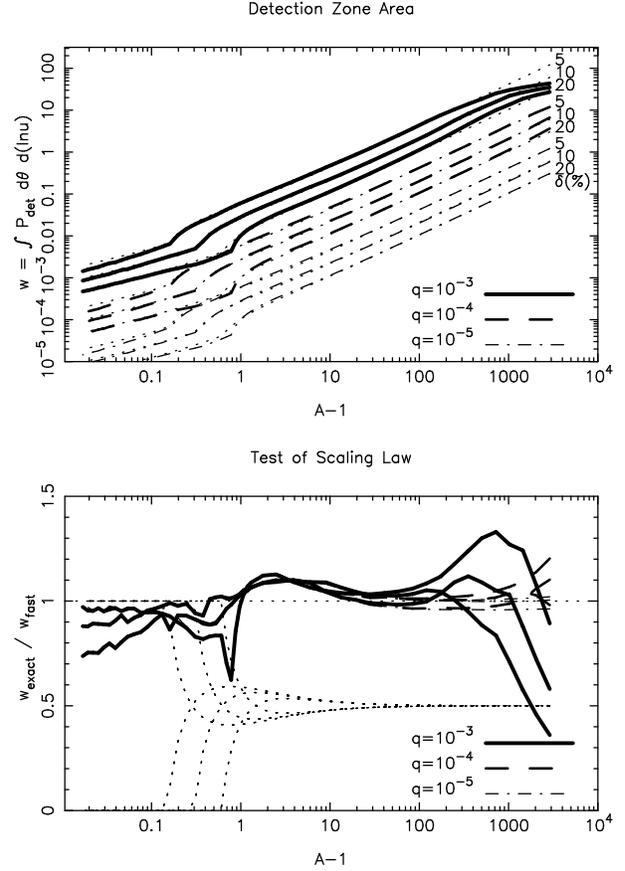}
\end{tabular}
\caption[] {\small
As in Fig.~{\ref{fig:scaling}}
but using the approximation in
defined in Eqn.~(\ref{eqn:wcut}) rather than Eqn.~(\ref{eqn:w}).
\label{fig:wcut}}
\end{figure}

At low-ish magnifications, $A<2$, Fig.~\ref{fig:scaling} shows that
the analytic formula over-predicts detection zone areas 
by factors of up to $\sim3$.
The structure is independent of $q$ but depends on $\delta$ and $A$,
due to the complicated structure of the detection zones as seen in 
Fig.~\ref{fig:zones}.
In this regime, $A\approx1-2\,u^{-4}$,
$F_+\approx u^{-2}$, $F_-\approx2\,u^{-2}$, and $F\approx3\,u^{-2}$.
With $F_-\approx2\,F_+$, the analytic formula
gives the minor image twice the detection area of the major image.
The $F_+$ formula has correct asymptotic behaviour at both high and low 
magnifications, so the problem is with the $F_-$ formula.
In fact at low magnifications the radial lobes of the minor image
merge and disappear, as seen in the right two columns of Fig.~\ref{fig:zones}.
This cuts the total detection zone area by a factor of about 3
when $A$ drops below a threshold,
\begin{equation}
	A -1 \approx \fracd{2}{u^4} \ltsim 3\,\delta
\ .
\end{equation}
As shown in Fig.~\ref{fig:wcut},
a fairly successful attempt to repair this deficit is
\begin{equation}
\label{eqn:wcut}
	w = \fracd{\pi\,q}{3\,\delta}
	\left( \fracd{A-1}{A+1} \right)^{1/2}
	\left( A + \left(1+A\right)\, C\left( x \right) \right)
\ ,
\end{equation}
where $x\approx3\,\delta/(A-1) \approx 3\,u^4\,\delta/2$, and
\begin{equation}
	C(x) \approx max\left[ 0, \fracd{ 1 - x^2 }{ 1 + x^2 } \right]
\end{equation}
cuts off the minor image contribution 
at the appropriate threshold.
However, this makes $F_-$ depend on $\delta$ as well as $A$.

\subsection{ Detection Zone Crossing Timescale }

\label{sec:crossing}
The cadence of observations aiming to detect planet-like anomalies
should ideally be matched to the
time it takes for the images to cross the detection zones, so
that each measurement probes for planets in an independent
region of the lens plane, rather than overlapping with the region 
already sampled by previous measurements.
This will clearly depend on the size of the detection zones, 
and on the speed at which the images move on the lens plane.

If we approximate the width of the detection zone as 
the square root of its area, then
the detection zone crossing timescale is
\begin{equation}
	t_\pm \approx \left( \fracd{w_\pm}{v^2} \right)^{1/2}
	\approx \left( \fracd{q\, F_\pm(A)}{\delta\,v^2}\right)^{1/2}
\ ,
\end{equation}
where $w_\pm=q\,F_\pm(A)/\delta$ is the detection zone area 
in the $\dd\theta\,\dd\ln{u}$ metric,
and $v$ is the corresponding image speed,
\begin{equation}
\label{eqn:vv}
	v^2 = 
	\left( \fracd{\dd \theta}{ \dd t} \right)^2
	+ 
	\left( \fracd{\dd \ln{u_\pm}}{\dd t} \right)^2
	=
	\dot{\theta}^2
	+
	\left( \fracd{\dot{u}_\pm}{u_\pm} \right)^2
\ .
\end{equation}
We show below that $v$ is actually the same for both images,
they move at the same speed
on the $\theta$ vs $\ln{u}$ plane.
The image speed and planet detection timescale are plotted
in Fig.~\ref{fig:afvt} for an event with $A_0=5$.

To evaluate the azimuthal velocity $\dot{\theta}$, note that
both images sweep around in $\theta$ at the same rate 
as the unlensed source position. 
With $x=(t-t_0)/t_E$ and $y=u_0$,
the unlensed source position moves with
\begin{equation}
	\dot{x}=\fracd{1}{t_E}
\ ,	\hspace{5mm}
	\dot{y}=0
\ .
\end{equation}
Differentiate $\theta=\arctan{\fracd{y}{x}}$ to find
\begin{equation}
\label{eqn:dthetadt}
	\dot{\theta} 
	= \dot{x}\, \fracd{\dd \theta}{\dd x}
	= \fracd{-\dot{x}\,y}{x^2+y^2}
	= \fracd{-u_0}{ u^2\, t_E}
\ .
\end{equation}

Evaluating the $\ln{u_\pm}$ image velocities is more involved, but 
leads to a simple result.
First, differentiate $u^2=x^2+y^2$ to find
\begin{equation}
	\dot{u} = \dot{x}\, \fracd{\dd u}{\dd x}
	= \fracd{\dot{x}\, x}{u} = \fracd{x}{u\, t_E}
\ .
\end{equation}
Next, differentiate
$\left(u_\pm\right)^2=\left(T\pm B\right)/2$
to find
\begin{equation}
\fracd{\dot{u}_\pm}{u_\pm}
	= \fracd{\dot{T}\pm\dot{B}
	}{ 4 \left( u_\pm \right)^2
	}
	= \fracd{ u\, \dot{u}\, \left(1\pm A\right)
	}{ 2 \left( u_\pm \right)^2
	}
\ ,
\end{equation}
where we have differentiated Eqn.~(\ref{eqn:tb}) to find
\begin{equation}
	\dot{T} = 2\, u\, \dot{u}
\ , \hspace{3mm}
	\dot{B} = 2\, u\, \dot{u}\, A
\ .
\end{equation}
Then, since $A=T/B$, and $\left(u_\pm\right)^2=\left(T\pm B\right)/2$,
\begin{equation}
\label{eqn:dlnuidt}
\fracd{\dot{u}_\pm}{u_\pm}
	= \fracd{u\, \dot{u}\, \left(B\pm T\right)}{B\, \left(T\pm B\right)}
	= -\fracd{u\, \dot{u}}{B}
	= - \fracd{x}{t_E\,B}
\ .
\end{equation}
Notice that both images have the same velocity in $\ln{u}$,
as well as in $\theta$, so that the image velocity $v$ is the same for both 
images.

Substituting Eqns.~(\ref{eqn:dlnuidt}) and (\ref{eqn:dthetadt})
into (\ref{eqn:vv}), the image speed on the $\theta,\ln{u}$ plane is
\begin{equation}
	\left( v\,t_E \right)^2
	= \fracd{u_0^2}{u^4} + \fracd{x^2}{B^2} 
	= \fracd{1 + 4\, u_0^2/u^4 }{u^2+4}
	= \fracd{u^4 + 4\, u_0^2 }{u^4\left( u^2+4 \right)}
\ .
\end{equation}
The image speed is plotted in Fig.~\ref{fig:afvt} for an event with $A_0=5$.
The azimuthal velocity $\dot{\theta}$ dominates near the peak, and the
radial velocity $\dd \ln{u}/\dd t$ dominates in the wings of the lightcurve.
For large $u^2$, the image speed varies
as $v\,\te \approx 1/u$.
The maximum speed $v_0\,\te=1/u_0$ is reached at the peak of the event,
at $t=t_0$, where $u=u_0$.

The planet anomaly crossing timescale is given by
\begin{equation}
\left( \fracd{t_\pm}{t_E} \right)^2
	= \fracd{w_\pm}{\left( v\, t_E \right)^2 }
	= \fracd{q\, F_\pm(A)}{\delta}
	\fracd{u^2+4}{1+4\,u_0^2/u^4 }
\ .
\end{equation}
Crossing times for the major and minor images are shown
in the lower panel of Fig.~\ref{fig:afvt}. 
The crossing time is slightly larger for the minor image,
by a factor $\left(F_-/F_+\right)^{1/2}=\left( \left(A+1\right)/A\right)^{1/2}$
(see Eqns.~(\ref{eqn:fplus}) and (\ref{eqn:fminus})).
The crossing time at first rises
due to the increasing size of the detection zone,
and then drops to a minimum at the peak,
where the maximum image velocity $v_0=1/\left(\te\,u_0\right)$
is reached.
This minimum crossing time, at the peak of the event, is
\begin{equation}
\begin{array}{rl}
t_\pm(0) & = \te\, u_0\,
	\left( 
	\fracd{q}{\delta} 
	\left( A_0+\fracd{1}{2}\mp\fracd{1}{2} \right)
	\right)^{1/2}
	\left( \fracd{A_0-1}{A_0+1} \right)^{1/4}
\\ \\ & \approx \te\, \left( \fracd{q}{\delta\, A_0} \right)^{1/2}
\ .
\end{array}
\end{equation}
Here the final approximation, using $F_\pm \approx A$
and $u \approx A^{-1}$,
holds to 15\% or better for $A_0\gtsim3$.

For a specific example,
consider the crossing time for a Jupiter-like planet with $q=10^{-3}$,
in a typical event with $\te=30$~d.
For a peak magnification $A_0=5$, as in 
Fig.~\ref{fig:afvt}, we have $u_0=0.2$,
$F_+=5\,\left(4/6\right)^{1/2}=4.1$
and $F_-=6\,\left(4/6\right)^{1/2}=4.9$.
For a good data point with $\sigma\left(\ln{A}\right)=1$\%, and a detection
threshold at $\Delta\chi^2=100$,
the smallest detectable planet-like anomaly deviates by 
$\delta=\sigma(\ln{A})\,\left(\Delta\chi^2\right)^{1/2}=0.1$.
The crossing time for the major image is
\begin{equation}
t_+(0)=30\times0.2\,
	\left( \fracd{0.001}{0.1} 5 \right)^{1/2}
	\left( \fracd{4}{6} \right)^{1/4} 
	= 1.2~{\rm d}
\ .
\end{equation}
For an Earth-mass planet and a typical lens mass
$\ml \approx 0.3~\msun$ the mass ratio is $q=10^{-5}$,
and the crossing time is
\begin{equation}
t_\pm(0)
	\approx 2.3~h 
	\left( \fracd{\te}{30~{\rm d}} \right)
	\left( \fracd{q}{10^{-5}} \right)^{1/2}
	\left( \fracd{\delta}{0.1} \right)^{-1/2}
	\left( \fracd{A}{10} \right)^{-1/2}
\ .
\end{equation}

\section{ Optimizing a Microlens Planet Search }
\label{sec:strategy}

\subsection{The Observer's Dilemma}
\typeout{dilemma}

In this section we consider how an observer might
try to optimize a microlens planet search.
We assume that the observer has many targets to choose from.
This is a good assumption because MOA~II and OGLE~III are finding
$\sim600-1000$ events each year.
The observing time available on each night during the 
winter months when the Galactic Bulge is visible from 
a southern hemisphere site is of order 10 hours.
A dedicated agile telescope spending 2 minutes per target 
could in principle visit over 100 targets per night.
However, why should equal time be devoted to all targets?
Surely the brighter and higher-magnification targets warrant
more attention.
By skipping fainter and/or weakly magnified sources, we can spend
more time on the more favourable ones.
The resulting planet detection zones, accumulated over all targets 
during the night, will then be larger, increasing the
chances of discovering a planet.
At the other extreme, when one source is very highly magnified, should we 
attend exclusively to that source, and skip all the others,
or should we reserve some time for a few of the other sources as well?
This is the microlens observer's perpetual dilemma.
The solution we propose is to observe always in a way that 
aims to maximise the probability of planet discovery.

\subsection{Accuracy of Photometry}
\typeout{accuracy of photometry}

Because detection zone areas scale as $w \propto \sigma^{-1}$,
e.g.\, Eqn.~(\ref{eqn:w}),
a critical issue is the accuracy of photometric measurements
that can be achieved, and the rate at which that accuracy
improves with exposure time.
We assume that the data analysis is close to optimal, so that photon 
counting statistics dominate the noise budget.
Thus CCD readout noise, cosmic ray hits, and other noise sources
are neglected in comparison with the Poisson noise from detected
star and sky photons.
The signal-to-noise ratio then increases as the square-root of the exposure 
time,
\begin{equation}
\label{eqn:sigma}
	\sigma(\ln{A}) = \left( \fracd{\tau}{\Delta t} \right)^{1/2}
\ ,
\end{equation}
\typeout{eqn:sigma}
where $\Delta t$
is the exposure time, and $\tau$ is the exposure time required to reach
a signal-to-noise ratio of 1.

The parameter $\tau$ controls the exposure time needed to 
obtain information on the current magnification $A$.
It depends on the telescope collecting area, the detector sensitivity and 
bandwidth, on the brightness of the 
magnified source star, and the degree of dilution of its photons by sky 
background and by other stars that are blended with it.
Including these three sources of Poisson noise, 
\begin{equation}
\label{eqn:tau}
\tau = \fracd{ \fstar + \fblend + \fsky }
	{ \fstar^2}
\ ,
\end{equation}
\typeout{eqn:tau}
where $\fstar$, $\fblend$ and $\fsky$ are the number of detected
photons per unit time from the magnified source star, 
from the lens star and other stars blended with the source star, 
and from the sky, respectively.
We elaborate these three Poisson noise sources below.

For a star of magnitude $m_\star$
and spectral energy distribution $f_\lambda(\lambda)$,
we observe thru the atmospere with transmission $T(\lambda)$,
with a detector effective area $A_{\rm eff}(\lambda)$.
The photon detection rate can be evaluated precisely as
\begin{equation}
\fstar = \int
	\fracd{ f_\lambda(\lambda)\, \dd\lambda }
		{ h\ \nu }
	A_{\rm eff}(\lambda)\
	T(\lambda)
\ ,
\end{equation}
or approximately as
\begin{equation}
\label{eqn:nstar}
\fstar	\approx \fvega(\lambda)\, 
	T(\lambda)\, 10^{- 0.4 m_\star }
\ .
\end{equation}
\typeout{eqn:nstar}
The photon detection rate from Vega (magnitude 0) is
\begin{equation}
\fvega(I)	\approx
	500\,{\rm s}^{-1}\,
	\left( \fracd{A_{\rm eff}}{\cmcm} \right)
	\left( \fracd{ \Delta\lambda }{ \rm \AA } \right)
\ ,
\end{equation}
for a telescope with mean effective area $A_{\rm eff}$
over a bandwidth $\Delta\lambda$
near the $I$ band, where most microlens observations are taken
(for the $V$ band, Vega's flux is 1000 rather than 500 
photons~cm$^{-2}$\AA$^{-1}$s$^{-1}$).

For the source star, magnified by a factor $A$, 
\begin{equation}
\fstar = \fsource\,A\, T
	= \fvega\, T\, 10^{-0.4\,m_\star}
\ .
\end{equation}

Stars blended with the magnified source star contribute Poisson noise to the 
measurement.
The source flux $\fsource$ and blend flux $\fblend$ 
are normally measured by fitting observed 
lightcurves (corrected for atmospheric transmission) with the model
\begin{equation}
	f(t) = \fsource\,A(t) + \fblend
\ .
\end{equation}
When using differential flux measurements $\Delta f(t)$, obtained by 
a difference image analysis, the reference flux added to these is
somewhat arbitrary. As a consequence, the blend flux $\fblend$ arising
from the lightcurve fit is also somewhat arbitrary, and
can even be negative in some cases.

The blend flux contributing to the Poisson noise includes not only
flux from the lens star, and any other stars that are 
``exactly'' coincident on the sky with the magnified source star,
but also stars that are close enough on the sky so that the
point-spread functions overlap.
If $m_i$ is the magnitude and $\theta_i$ 
is the angular separation of star $i$ from the target star,
the blend flux contributing Poisson noise to the measurement is
\begin{equation}
\fblend(\Delta)
= \fstar \sum_i 10^{-0.4\ \left( m_i - m_\star \right)}\
	e^{-\left( \theta_i/\Delta \right)^2}
\ .
\end{equation}
This expression assumes a gaussian point-spread function
with standard deviation $\Delta$, and optimal extraction 
to measure the target star flux.
Note that the Poisson noise due to blended stars increases with the seeing.
Although it is not yet done in practice,
the specific dependence on seeing for each microlens target can be 
evaluated in advance from a good-seeing image of the starfield,
for example the OGLE or MOA finding-chart images made available for each 
event.

Finally, the detection rate of sky background photons overlapping with the 
target star is
\begin{equation}
\label{eqn:nsky}
	\fsky = \fvega\ \Delta\theta^2\ 10^{-0.4\,\musky}
\ ,
\end{equation}
where $\musky$ is the magnitude of a square arcsecond of sky,
and $\Delta\theta^2$ is the solid angle subtended by the photometric 
aperture (for aperture photometry) or by the 
point-spread function of the star 
images (for psf-fitting photometry) in square arcseconds.
The sky brightness, including e.g.\ airglow, zodiacal light and
scattered moonlight,
may be evaluated using a sky model, e.g. \cite{ks91,p03}.
The effective sky coverage of
a gaussian point-spread function is
\begin{equation}
\Delta \theta^2 = 4\,\pi\,\Delta^2
	= \fracd{\pi }{2\,\ln{2}}\ W^2
\ ,
\end{equation}
where $\Delta$ is the standard deviation and $W$ is the full-width at half-maximum 
(FWHM) of the gaussian point-spread function.
Atmospheric seeing is usually reported in terms of $W$.

Combining the above equations, we can rewrite Eqn.~(\ref{eqn:tau}) as
\begin{equation}
\tau = \taustar + \taublend + \tausky
\ ,
\end{equation}
with
\begin{equation}
\taustar =
	\fracd{ 10^{0.4\,\magstar-9.7} }
	{ T(\lambda)\,
	\left( \fracd{\aeff}{\msq} \right)
	\left( \fracd{\Delta\lambda}{10^3{\rm \AA}} \right)
	}
\ ,
\end{equation}
\begin{equation}
\taublend = \taustar\,
	\sum_i 10^{-0.4\left( m_i - \magstar \right)}
	e^{-\left(\theta_i/\Delta\right)^2}
\ ,
\end{equation}
and
\begin{equation}
\tausky = \taustar\,
	\fracd{ 4\,\pi\,\Delta^2}{ T(\lambda) }
		10^{-0.4\left( \musky - \magstar \right)}
\ .
\end{equation}
These expressions make explicit how
$\tau$ depends on the magnified source star brightness
(magnitude $\magstar$),
on the nearby stars blended with the target 
(magnitude $m_i$, separation $\theta_i$),
on capabilities of the telescope 
(effective area $A_{\rm eff}$, bandwidth $\Delta\lambda$),
and on observing conditions
(sky brightness $\musky$, seeing $\Delta$, atmospheric 
transmission $T$).
When the sky and blend fluxes are negligible,
a 100s exposure with $\aeff=1~\msq$ and $\Delta\lambda=10^3$~\AA\
reaches 1\% accuracy at magnitude $\magstar=16.8$.

\subsection{ Optimal Exposure Times }

The ``worth'' of an observation,
from the perspective of planet hunting,
is proportional to the area of the resulting detection zone.
Combining Eqns.~(\ref{eqn:w}) and (\ref{eqn:sigma}), we see that
detection zone areas increase with
the square-root of the exposure time,
\begin{equation}
	w 
	= \fracd{q\,F(A)}{\delta}
	= q\,F(A)\, \left( \fracd{ \Delta t}{\tau\, \Delta\chi^2} \right)^{1/2}
	\equiv g\ \left( \Delta t \right)^{1/2}
\ .
\end{equation}
The proportionality constant $g$ characterizes
the ``goodness'' of observing this particular target,
\begin{equation}
\label{eqn:goodness}
g \equiv \fracd{ q\ F(A) }{ \left( \tau\ \Delta \chi^2 \right)^{1/2} }
\ .
\end{equation}
These $g$ values can be used to prioritise the events
that are available at any given time. They depend on the 
properties of the event, characteristics of 
the telescope, and on the present observing conditions.

The key point to note here is that the fractional measurement error
decreases with the exposure time,
$\sigma \propto \Delta t^{-1/2}$,
and this expands the detection zone area as $w \propto \Delta t^{1/2}$.
The detection zone area grows most rapidly
at the beginning of the exposure, with diminishing 
returns as the exposure progresses.
For this reason at some point it becomes advantageous
to abandon observations of this target in favour of moving on
to a fresh target that has not yet been observed.

Suppose that we are contemplating making observations of
$N$ targets during an upcoming night in which we expect to
have available a total observing time $t$.
How much exposure should we devote to each target?
For each target $i$ we can calculate the goodness factor $g_i$.
If we observe target $i$ with exposure time $\Delta t_i$,
then
\begin{equation}
	t = \sum_{i=1}^N \Delta t_i
\end{equation}
is the total observing time. The total worth 
of observing the $N$ targets is
\begin{equation}
\label{eqn:wn}
	W_N = \sum_{i=1}^N g_i\, \left( \Delta t_i \right)^{1/2}
\ .
\end{equation}

\begin{figure}
\begin{tabular}{c}
\psfig{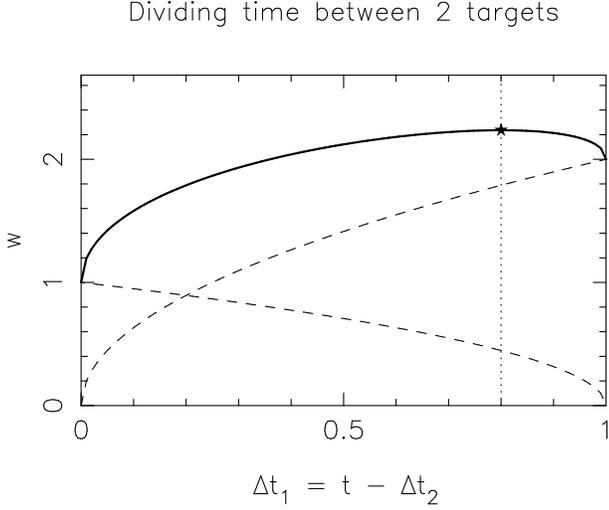}
\end{tabular}
\caption[] {\small
When observing $N$ microlens events,
for which the planet detection zone areas $w_i$ grow with
exposure time $\Delta t_i$ as $w_i = g_i\, {\Delta t_i}^{1/2}$,
the total detection zone area $W=\sum_i w_i$
is maximised when the
available exposure time is divided in proportion to $g_i^2$.
Optimisation for the case $N=2$ is illustrated here.
\label{fig:two}}
\end{figure}

Given a fixed total observing time $t$,
we can optimize the exposure times
by solving $\partial W_N / \partial \Delta t_i = 0$.
For example, with $N=2$ targets, as illustrated in 
Fig.~\ref{fig:two},
the total time is $t = \Delta t_1 + \Delta t_2$, and
the sum of the detection zone areas is
\begin{equation}
	W_N = g_1\,\left(\Delta t_1\right)^{1/2}
	+ g_2\, \left( t - \Delta t_1 \right)^{1/2}
\ .
\end{equation}
Maximizing $W_N$ gives
\begin{equation}
	0 = \fracd{\partial W_N}{\partial \Delta t_1}
	= \fracd{ g_1}{2\,\left(\Delta t_1\right)^{1/2}}
	- \fracd{ g_2}{2\,\left(\Delta t_2\right)^{1/2}}
\ ,
\end{equation}
and thus $\Delta t_1 / \Delta t_2 = \left( g_1 / g_2 \right)^2$.
Similarly, for the general case of $N$ targets, the optimal exposure times 
that maximize $W_N$ also satisfy $\Delta t_i \propto g_i^2$,
and are therefore given by
\begin{equation}
\label{eqn:dtopt}
	\Delta t_i = \left( \fracd{ g_i}{ G_N } \right)^2\,t
\ ,
\end{equation}
where
\begin{equation}
	{G_N}^2 \equiv \sum_{i=1}^N g_i ^2
\ .
\end{equation}
If we adopt the optimal exposure times, 
substituting Eqn.~(\ref{eqn:dtopt}) into (\ref{eqn:wn})
gives the total worth of the observations as
\begin{equation}
        W_N = G_N\ t^{1/2}
\ .
\end{equation}

This analysis suggests that the optimal strategy 
to maximize the planet detection capability
is to observe all available targets, 
spending more time on the best targets,
using exposure times proportional to the square of the goodness,
$\Delta t_i \propto g_i^2$.
The optimal observer skips no targets.
As we will see, however, this conclusion is altered
when we take account of observing overheads.

\subsection{ Effect of Overheads}

\begin{figure}
\begin{tabular}{c}
\psfig{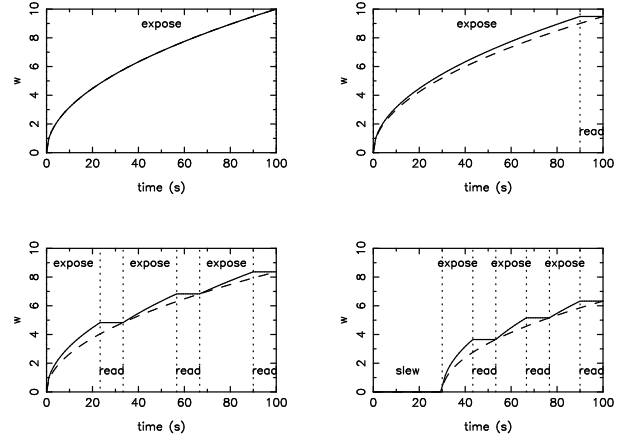}
\end{tabular}
\caption[] {\small
Illustration of the reduction in the net exposure time
and the corresponding degradation in planet hunting capability
for a 100s observation accounting for
a CCD readout time $\tread = 10$s,
for splitting the exposure into 3 sub-exposures
to avoid saturation, and 
for a telescope slew time $\tslew=30s$.
\label{fig:plotw}}
\end{figure}

In practice the CCD camera takes a finite time $\tread$ to read out,
and the telescope takes a finite time $\tslew$ to slew from one target
and settle into position on the next.
Typical readout and slew times are $\tread\sim10-20$~s
and $\tslew\sim1-3$~min.
To avoid CCD saturation, a single long exposure
may need to be broken up into a series of $n$ shorter exposures.
These overheads reduce the on-target exposure time
accumulated during an observation time $t$ to
\begin{equation}
        \Delta t = t - \tslew - n\,\tread
\ .
\end{equation}
These overheads diminish the planet hunting capability of
the observations, as illustrated in Fig.~\ref{fig:plotw}.
We must allow for these overheads when implementing
an optimal observing strategy.

If the CCD exposure is too long, the target will saturate.
If the CCD exposure is too short, the readout noise will
dominate over sky noise and information will be lost.
These considerations set the range that should be
considered for the CCD exposure time: 
\begin{equation}
	\tmin < \texp < \tmax
\ .
\end{equation}
A total exposure longer than $\tmax$ is accumulated by
taking a series of $n$ shorter exposures,
where
\begin{equation}
	\fracd{t - \tslew}{\tmax + \tread}
	< n <
	\fracd{t - \tslew}{\tmin + \tread}
\ .
\end{equation}
For bright targets where $\tmax$, the longest exposure that
avoids saturation, is less than $\tmin$, the shortest exposure
that avoids readout noise domination, the need to
avoid saturation must take precedence. 
Having $n>1$ protects against cosmic ray hits.
Once $n$ is decided, the duration of each exposure is
\begin{equation}
	\texp = (t - \tslew) / n - \tread
\ .
\end{equation}
Allowing for these overheads, the detection zone area becomes
\begin{equation}
\label{eqn:overheads}
	w = g\ \left( \fracd{\texp}{\texp + \tread} \right)^{1/2}
	\left( t - \tslew \right)^{1/2}
\ .
\end{equation}
The first bracket accounts for the reduction in on-target observing
time due to the CCD readout time.
We can absorb this term into the definition of $g$,
\begin{equation}
	g \rightarrow \fracd{g}
	{ \left( 1 + \left( \tread / \texp \right) \right)^{1/2}}
\ ,
\end{equation}
as shown by the dashed curves in Fig.~\ref{fig:plotw}.
The effect is to suppress interest in observing
targets that are so bright as to require inefficient
observations with $\texp < \tread$.
A target too bright for efficient observations with a large telescope
may thus remain a prime target for smaller telescopes.
In this way the scheme may serve well to coordinate
observations by a community with a variety of telescope types.

The second bracket in Eqn.~(\ref{eqn:overheads}), allowing for the slew 
time,
delays the onset of detection zone growth while the
telescope is moving from one target to the next.
We will see below that this term dictates
which of the less promising targets
to omit from the observing schedule.

\subsection{ Dividing Time among $N$ Events }

\begin{figure}
\begin{tabular}{c}
\psfig{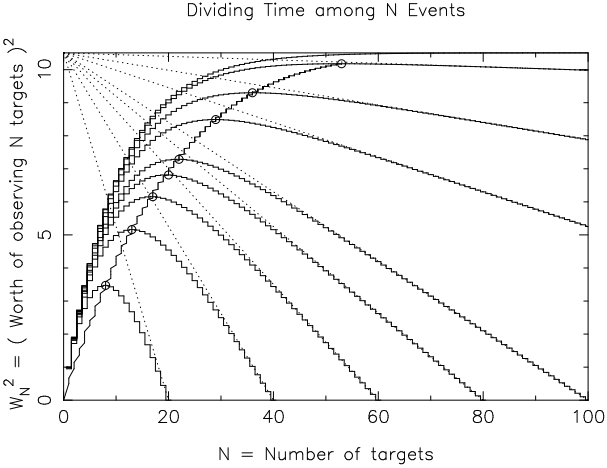}
\end{tabular}
\caption[] {\small
The number of microlens events to observe
is taken to maximise the total worth of the observations.
With too many targets, observing time is reduced
by the time required to slew the telescope from target to target.
This optimisation of $N$ is illustrated for 100 targets
with an exponential distribution of goodnesses, 
and for different total amounts of telescope time.
With less time fewer events are observed, and the 
total worth is reduced.
\label{fig:plopfake}}
\end{figure}

If we try to observe too many of the ongoing events,
we will spend all night slewing from target to target
and no time at all collecting photons from the targets.
If the total time available for observations is $t$,
and slew time is $\tslew$, then
the maximum number of targets we can contemplate observing is
\begin{equation}
\nmax = t / \tslew
\ .
\end{equation}

If we observe $N \leq \nmax$ targets,
the total worth of the observations will be
\begin{equation}
	W_N = G_N\,\left( t - N\,\tslew \right)^{1/2}
\ .
\end{equation}
We would like to maximize $W_N$.
The first term $G_N$ increases with $N$, and 
the second term $\left( t- N\,\tslew \right)^{1/2}$ decreases with $N$.
Therefore
$W_N$ has a maximum value for some $N < \nmax$.
This is the number of targets that we should observe to
maximize the planet hunting capability of our observations.

To make $W_N$ grow as fast as possible, 
sort the targets and consider them in order of decreasing goodness,
$g_N \geq g_{N+1}$.
We should keep target $N+1$ only if $W_{N+1} > W_N$.
To decide whether or not to retain target $N+1$, note that
\begin{equation}
\left(\fracd{W_{N+1}}{W_N}\right)^2
	= \left( 1 + \left( \fracd{g_{N+1}}{G_N}\right)^2 \right)
	\left( 1 - \fracd{\tslew}{t-N\,\tslew}\right)
\ .
\end{equation}
Target $N+1$ survives only if
\begin{equation}
g_{N+1} > 
\fracd{G_N\,\tslew^{1/2}}{\left( t - N\,\tslew \right)^{1/2}}
= \fracd{G_N}{\left( \nmax - N \right)^{1/2}}
\ ,
\end{equation}
where we have used $\nmax = t/\tslew$.

To illustrate this optimisation,
Fig.~\ref{fig:plopfake} shows the result of dividing time among $N=100$
targets with an exponential distribution of goodnesses $g_i$,
for several different total available observing times $t$,
corresponding to $\nmax=20, 40, ...$.
As available time $t$ increases, the optimal strategy spends more time on
each target, and also extends time to additional 
lower-priority targets.

\begin{figure*}
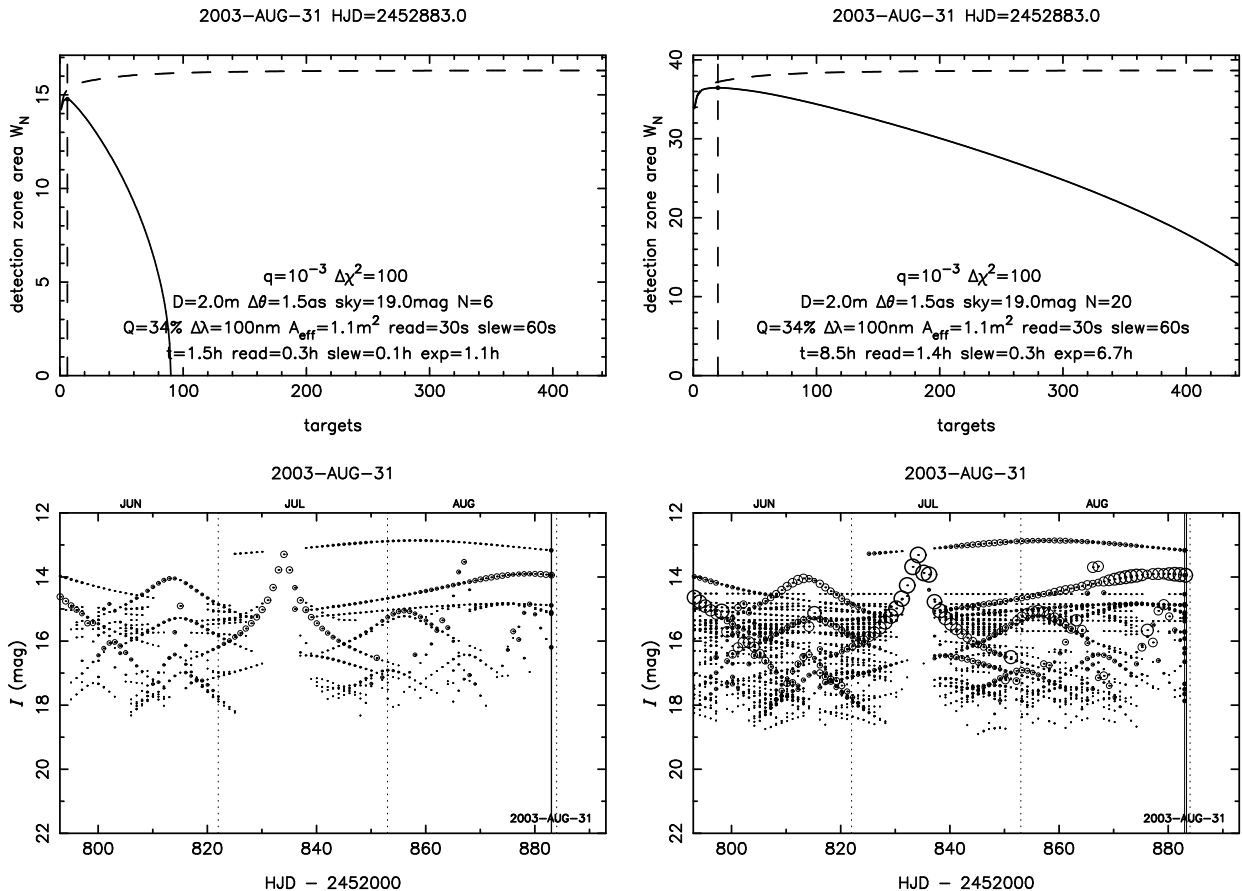

\begin{tabular}{cc}
\psfig{file=horne_fig12a.ps,angle=0,width=8cm}
&
\psfig{file=horne_fig12b.ps,angle=0,width=8cm}
\end{tabular}
\caption[] {\small
Optimal sampling of 443 OGLE events
available on 2003~Aug~31.
Exposure times are chosen to
maximize the total planet detection zone area,
for a 2~m telescope with fixed total observing time 
$1.5$h (left) and $8.5$h (right) per night.
Observing too many targets is inefficient because of the
120s telescope slew time.
Observing too few targets is inefficient because
planet detection zone areas grow only as $t^{1/2}$.
On the resulting lightcurves (lower panels), the plot symbol
areas are proportional to the allocated exposure time.
On most nights observing time spreads over many targets.
On some nights one high-magnificaiton target
captures most or all of the attention.
\label{fig:plopreal}}
\end{figure*}

To further illustrate, more realistically,
we consider in Fig.~\ref{fig:plopreal}
the recommended observations from among 443 OGLE 
events that were available on 2003~Aug~31.
To decide on the observing strategy,
we first fit a PSPL lightcurve model to the OGLE 
data on each event to evaluate the event parameters.
This results in predicted magnitudes and magnifications for
each target on each night in question.
We next evaluate the goodness factors $g_i$
for a telescope with effective area $A_{\rm eff}=1.1$m$^2$,
with a sky magnitude 19.
We assume a slew time $\tslew=60$s, a readout time $\tread=10$s,
a maximum exposure time $\texp<600$s.

 In the top panels of Fig.~\ref{fig:plopreal},
the dashed curve shows how $W_N$ would increase monotonically 
if there were no slew time.
The solid curve shows show how $W_N$ at first increases 
with $N$ but then decreases as slew time becomes important.
For $t=1.5$h (left panel of Fig.~\ref{fig:plopreal}), 
the maximum number of targets that could be observed is 
$\nmax=t/\tslew=90$, but the optimal sampling to maximise
$W_N$ undertakes observations of just $N=6$.
For $t=8.5$h (right panel) all 443 targets can be observed, but the maximum 
of $W_N$ occurs at $N=20$.

 In the bottom panels of Fig.~\ref{fig:plopreal},
the resulting lightcurves are shown when this strategy is employed on every 
night.  The area of the plot symbols are proportional to the observing time 
allocated to each target. On most nights the optimal sampling spreads 
observing time over many targets. On a few nights when one very high 
magnification event is available, that target captures most or all of the 
recommended observing time.  One bright target receives some attention even 
though its magnification is small. The observations include fainter targets
when more observing time is available.  Targets fainter than the sky 
are seldom scheduled.

\section{Discussion}
\label{sec:discussion}

\subsection{ Detection is not Characterisation }

We must emphasize that the scheme outlined above
is designed to detect anomalies, not to characterise them.
Observing many targets for the recommended exposure time
can be advocated only so
long as each new observation indicates that no
significant anomaly is underway.
It is therefore best if a rapid reduction of each new observation
can be undertaken with sufficient accuracy and reliability 
to check each new data point for consistency or otherwise
with the PSPL model.
This is feasible because only one or at most a few stars on
each CCD image will be undergoing microlensing at a given time.
The sub-image around the target of interest can be quickly reduced to 
measure its brightness with respect to nearby comparison stars.
In practice the real-time image-subtraction
pipelines currently in use by PLANET and RoboNet
can reduce each CCD image within a few minutes
of the end of the exposure.

Whenever a significant anomaly is identified,
the observer can temporarily suspend the anomaly-hunting
strategy of observing many targets in sequence,
returning to the target that offered up the anomalous data point.
Additional observations of this target then aim to establish either
that an anomaly is in progress, or else to dismiss the false
alarm caused by unreliable data.
If the return observations fail to confirm the anomaly,
then the anomaly-hunting observations can resume.
If the return observations confirm the anomaly, then continuous
observations are initiated to clarify the nature of the anomaly,
and an alert can be issued to trigger follow-up observations
on other available telescopes.
An implementation of this is the SIGNALMEN
anomaly detector \cite{d+07}.

By following this two-stage approach -- prioritised multi-target
anomaly hunting punctuated by episodes of 
anomaly confirmation and characterisation --
we can simultaneously maximize the opportunity to detect
anomalies by observing a large number of targets,
while retaining the ability to reliably establish
the nature of anomalies that we detect.
If the second-stage is omitted from the observing strategy,
the risk is a series of single-point anomalies will be found
whose identity cannot be securely established.

\subsection{ Dynamic Priorities }
\label{sec:priority}

\begin{figure}
\begin{tabular}{c}
\psfig{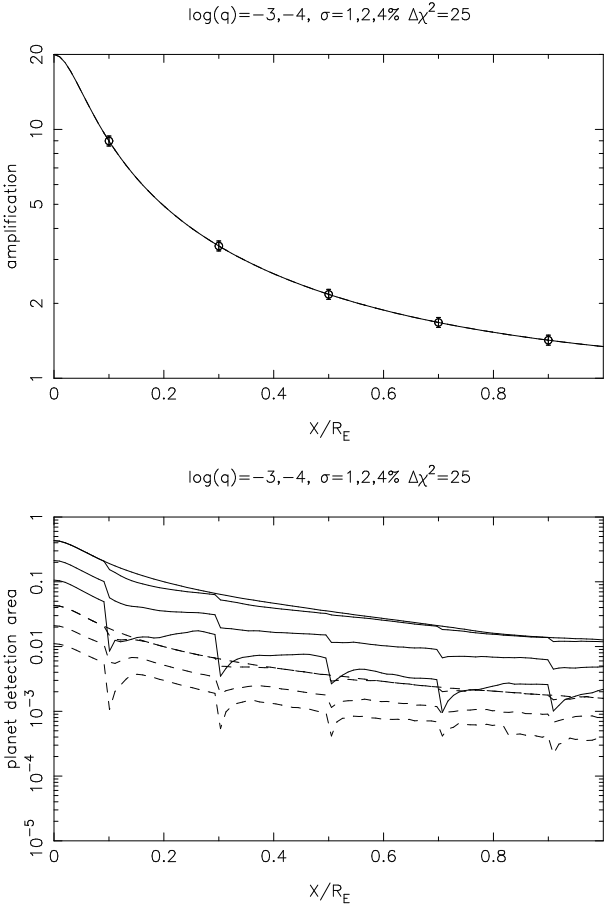}
\end{tabular}
\caption[] {\small
Top panel shows 5 data points at times $t_i$
with accuracy $\sigma_i=5$\% on the decline
of a lensing lightcurve with maximum magnification $A_0=20$.
Bottom panel shows the evolving priority
given to proposed new data points with accuracy 
$\sigma=1$, 2, and 4\% (top to bottom)
when searching for planets with mass ratio
$q=10^{-3}$ (solid curves) and $10^{-4}$ (dashed curves).
The priority for an isolated data point is proportional to
the planet detection zone area,
$\Omega \propto q (2A(t)-1) / \sigma$,
where $A(t)$ the time-dependent magnification.
The priority drops when the new observation
would probe for planets inside a detection zone
already established by previous observations.
The priority then recovers on the timescale needed
for images to cross the detection zones.
\label{fig:fivepoints}}
\end{figure}

When event parameters change significantly during a night,
or when the image positions fall inside detections zones
from previous observations, then the results derived in
the previous section are no longer strictly valid.
Movement of the images will increase, while overlap 
will decrease the detection zone areas.
We are nevertheless hopeful that the optimisation scheme
advocated above will still be helpful in guiding
follow-up observations.

One way to cope with the more general situation is
to employ a scheme with continually-evolving target priorities.
The highest-priority target is observed.
The priority of that target must then fall dramatically,
since immediate re-observation would probe
for planets inside the detection zone just carved out.
The priority should then recover in due course,
as changes in the event geometry move the image position
outside of the detection zone.
Such a scheme may be ideal for fully automatic
follow-up observations with robotic telescopes,
but could also be used by human observers
willing to follow directions from a computer programme.

A dynamical priority scheme of this sort is
illustrated in Fig.~\ref{fig:fivepoints}.
The top panel of  Fig.~\ref{fig:fivepoints} shows
a set of data points with accuracy $\sigma_i=5$\%
at times $t_i$ during the decline of an event
with peak magnification $A_0=20$.
The lower panel shows the evolving priority
given to a proposed new data point at time $t$
with accuracy $\sigma=1$, 2, and 4\%
when searching for planets with mass ratio
$q=10^{-3}$ and $10^{-4}$.
The priority is evaluated numerically as
the increase in detection zone area arising
from the proposed new data point.

The dips in priority evident in the lower panel of 
Fig.~\ref{fig:fivepoints}
indicate the reduced planet hunting capability caused
by the overlap of detection zones when the
new data point probes for planets inside the
detection zone of an earlier measurement.
We see in Fig.~\ref{fig:fivepoints} that
the reduction is small for $\sigma=1$\%
and substantial for $\sigma=4$\%.
This is because the old $\sigma=5$\% data are important
when the new data point is of similar accuracy, but unimportant
when the new data point has much higher accuracy.
We see also in Fig.~\ref{fig:fivepoints} that the priority
recovery time is faster for $q=10^{-4}$ than for $10^{-3}$.
This is plausible since detection zone sizes scale as $q^{1/2}$.
The rather irregular recovery arises from the complicated shapes
of the detection zones (Fig.~\ref{fig:zones}).

In practise there will be not just a single previous measurement
with accuracy $\sigma$, but rather a set of prior measurements
at times $t_i$ with accuracies $\sigma_i$.
Noting that independent measurements combine optimally with 
$1/\sigma^2$ weights, the net effect at time $t$ of all prior measurements
may be approximated by using the scheme
\begin{equation}
\fracd{1}{\sigma^2(t)} =
	\sum_i \fracd{ M\left[ \left( t - t_i \right) / s_i \right]
		}{\sigma^2_i}
\ ,
\end{equation}
where $s_i$ is an ``expiration time'' for the observation at time $t_i$,
and $M(x)$ is a ``memory function'', 1 for $t=t_i$
and decreasing to 0 for $t>>t_i$, effectively forgetting sufficiently
old observations.
Possibilities for the memory function are Gaussian or Lorentzian:
\begin{equation}
	M(x) = e^{-x^2/2}
\ , \hspace{5mm}
	M(x) = \fracd{1}{1+x^2}
\ .
\end{equation}
The detection zone crossing, time worked out in Section.~\ref{sec:crossing},
provides a suitable expiration time $s_i$.

The new detection zone area grows more slowly due to overlap
with earlier zones.  It is as if an exposure time $\tdone$
has already been done to achieve the accuracy
$\sigma(t) = \left(\tau/\tdone\right)^{1/2}$.
The new exposure time $\Delta t$ then adds to $\tdone$, increasing
the detection zone area by
\begin{equation}
\Delta w = g \left( \left(\Delta t + \tdone \right)^{1/2} - \tdone^{1/2} 
		\right)
\ .
\end{equation}
with
\begin{equation}
	\tdone = \fracd{\tau}{\sigma^2(t)}
	= \tau\ \sum_i
	\fracd{ M\left[ \left( t - t_i \right) / s_i \right]
		}{ \sigma^2_i }
\ .
\end{equation}

Another relevant consideration is that slew times are not equal for all 
targets.
The slew time is zero for the current target, and 
for other targets should increase with 
their angular distance from the current target.
If we include the slew time, then the increase in detection zone area is
\begin{equation}
\Delta w = g \left( \left(
	{\rm max}\left[ 0, \Delta t - \tslew \right]
	+ \tdone \right)^{1/2}
		- \tdone^{1/2} 
		\right)
\ .
\end{equation}
If we require the exposure time to be not shorter than some minimum time,
perhaps some multiple of the CCD readout time, in order to have a 
reasonably high observing efficiency, then one compares the options of
observing longer on the present target without slew time vs 
slewing to another target.
As $\tdone$ increases on the current target, the
potential for increasing its detection zone area declines until it
becomes better to slew to and expose on the next target.

\begin{figure}
\begin{tabular}{c}
\psfig{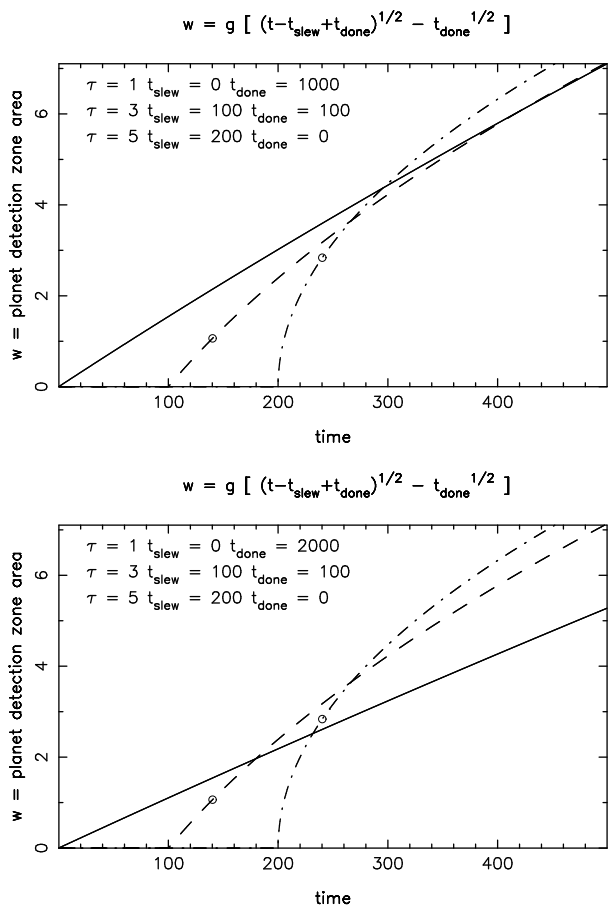}
\end{tabular}
\caption[] {\small
The slew time is zero for the current target, but
significant for two others. Observations of the current
target should continue until the marginal improvement in
detection zone area becomes less for this target than for
one of the alternatives, taking the slew times into account.
\label{fig:whentoslew}}
\end{figure}

This scenario is illustrated in Fig.~\ref{fig:whentoslew}.
Here three targets are considered.
We are currently exposing on target 1, with slew times of 100s and 200s
to reach targets 2 and 3.
We consider a minimum exposure time of 40s. The circles show the result
of slewing to an alternative target and exposing for the minimum time.
In the top panel we have accumulated a 1000s exposure on target 1.
The circles for both alternative targets are below the solid curve, so we 
should not slew.
In the bottom panel we have accumulated a 2000s exposure on target 1,
and this increase in $\tdone$ reduces the slope of the solid curve
to such an extent that the circle on target 3 is now just above it.
At this point we should therefore decide to slew and expose on target 3,
rather than remaining on target 1.
This cycle may be iterated throughout the night.

 We expect an optimisation scheme based on approximations 
like those described above to be helpful in deciding how long 
to continue observing the 
present target, and which target is the best one to observe next. 
We have not yet simulated this possibility in great detail, but 
outline the concept here as a possible starting point
for a self-organising scheme that may be suitable 
for coordinating optimal microlens observations by 
a heterogeneous network of telescopes.
Assuming rapid sharing of information among the telescope nodes,
each telescope can independently decide which target to observe next,
taking into account prior observations made by all other telescopes,
with their various times and accuracies.

\subsection{ Uncertain Event Parameters}
\label{sec:eventerrors}

The event parameters $t_0$, $\te$, and $A_0$ are often
uncertain and correlated in the early stages of
an event before the observations have sampled 
both sides of the lightcurve peak.
With highly uncertain event parameters, 
large errors may arise in the assigned target priorities.
A frequent example occurs when an early fit to the rising
part of the lightcurve suggests a very high magnification event
that later turns out to be of only modest magnification.
How will such uncertainties affect our strategy?

One happy aspect: the detection zone areas depend
on current values of the magnification $A$ and star magnitude $m_\star$,
rather than on the values at the peak of the lensing event.
This is helpful in the early stages of an event when the
eventual peak magnification is still difficult to predict.
On the other hand, in a highly-blended event the true magnification
of the source star can be higher than the apparent magnification.

The event parameter uncertainties remaining after fitting the PSPL
lightcurve model to the extant data points can be quantified, for 
example by using the parameter covariance matrix or 
Markov-Chain Monte-Carlo techniques.
The corresponding uncertainy in the event priority may then be taken into 
account using a Bayesian average over the posterior probability 
distributions.

One may also contemplate giving priority to observations aiming to reduce
uncertainty in the event parameters. This secondary goal will then need to
be traded-off in some satisfactory way with the primary goal of discovering
planets. An anomaly found on the rise should attract attention, making it
likely that accurate event parameters will be nailed down by observations
across and after the peak. For an event well past the peak, it may be too
late for additional observations to pin down uncertain event parameters, or
additional observations at critical stages may help a lot to break the
ambiguity between blending, magnification, and event timescale.  Targets
could be given reduced priority when their event parameters are uncertain
and there is little prospect of improving them, or higher priority when a
critical observation would help to nail down the uncertain prameters. This
issue needs careful investigation. 

\begin{figure}
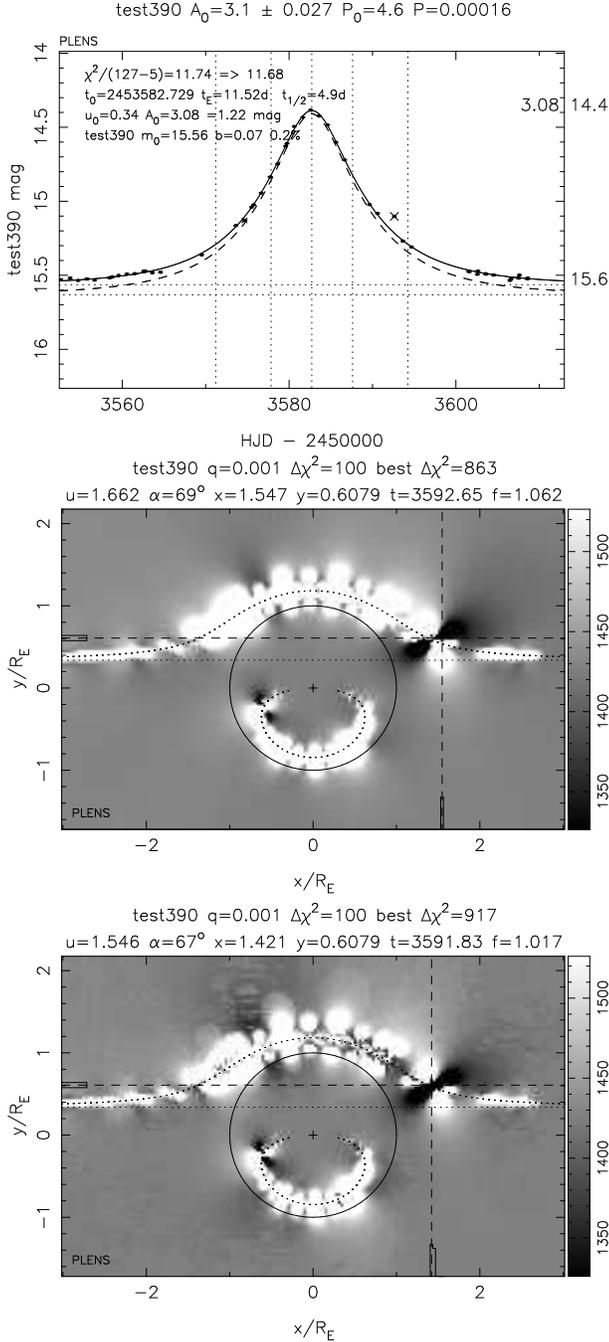

\begin{tabular}{c}
\psfig{file=horne_fig15a.ps,angle=270,width=8cm}
\\
\psfig{file=horne_fig15b.ps,angle=270,width=8cm}
\\
\psfig{file=horne_fig15c.ps,angle=270,width=8cm}
\end{tabular}
\caption[] {\small
Top panel: OGLE III observations of OGLE-2005-BLG-390.
One data point occurs during a planet anomaly.
The point-source point-lens (PSPL) model fits 5 parameters,
the peak time $t_0$, peak magnification $A_0$, 
Einstein radius crossing time $\te$,
the source flux $\fsource$ and blend flux $\fblend$.
Middle panel: Greyscale representation of $\chi^2(x,y)$, moving a
planet with mass ratio $q=10^{-3}$ on the $(x,y)$ lens plane,
holding fixed the 5 PSPL parameters.
The $\chi^2$ increases by 100 or more in the white areas,
where the planet is ruled out, and decreases by 100 or more in the
black areas.
Bottom panel: The $\chi^2(x,y)$ map re-fitting the 5 PSPL parameters
for each planet position, showing the smaller size of the resulting
planet detection zones.
\label{fig:test390}}
\end{figure}

Our estimates of detection zone areas assume that the underlying lens
parameters are or will be well constrained by observations outside the
planet anomaly.  When this is not the case, then the actual detection zones
will be smaller, because the loose event parameters can shift the model
toward the anomalous data points that would otherwise be able to detect 
or rule out
planets. Fig.~\ref{fig:test390} illustrates this effect, where the
reduction of detection zone areas is considered for the OGLE data
on OGLE-2005-BLG-390. In this event, one OGLE data point occurs during a 
planet anomaly. The reduction of detection zone areas is noticeable but
not large enough in cases such as this to be a serious problem for our 
optimisation scheme. It would be a more serious problem for events
with only a few measurements covering the 
magnified part of the lightcurve.

\subsection{ Targeting Specific Types of Stars and Planets}
\label{sec:finetune}

As our knowledge of the exo-planet distribution function accumulates,
one might contemplate introducing a prior on the 
parameters $q$, $M_\star$, and $a$ in order to
target the planet search toward particular types of 
stars or planets.
For example, since
$\te \propto M_\star^{1/2}$,
fast events correspond on average to lower-mass stars.
Similarly, since $a \propto \te u_\pm$,
larger orbits can be targeted by observing slower events and 
observing longer after the event peak.
It is straightforward to tilt the search toward any specific parts
of parameter space. 
However, at this stage our knowledge of the cool planet distribution
is so scant that it is probably premature to invest
much effort into such fine-tuning.

\section{ Summary}
\label{sec:summary}

OGLE~III and MOA~II are discovering 600-1000 Galactic Bulge
microlens events each year.  
This stretches the resources available for intensive follow-up
monitoring of the lightcurves in search of planets near the
lens stars.
We advocate optimizing microlens planet searches
by using an automatic prioritization algorithm
based on the planet detection zone area probed by each
new data point.
We evaluate detection zone areas numerically 
and validate a plausible scaling law useful 
for rough but rapid calculations.
The proposed optimization scheme takes account of
the telescope and detector characteristics,
CCD saturation, readout time, and telescope slew time,
sky brightness and seeing,
past observations of microlensing events underway,
and the time available for observing on each night.
The current brightness and magnification of each target
are estimated by extrapolating fits to previous data points.
The optimal observing strategy then provides
a recommendation of which targets to
observe and which to skip, and a recommended exposure
time for each target, designed to maximize the planet
detection capability of the observations.
This must be coupled with rapid data reduction to
trigger continuous follow-up observations
whenever an anomaly is detected.
It is hoped that the algorithm will provide helpful guidance
to follow-up observing teams, and may be a useful 
starting point for optimising fully-robotic
microlens planet searches.

\subsection{WEB-PLOP}

An implementation of this optimisation scheme, Planet Lens
OPtimisation (PLOP or web-PLOP),  can be found at 
{\tt http://www.artemis-uk.org/web-PLOP/} \cite{s+08}.
 This system was designed with two motivations: to
provide an optimal target list for the automated observing of the
RoboNet project \cite{b+07,t+09},
and also to provide such lists
to human observers at any telescope. It is formed of two parts.
First, a user interface web form takes input for the telescope and
observing conditions parameters ($A_{\rm eff}$, $t_{\rm slew}$, $t_
{\rm read}$, $\mu_{\rm sky}$, $\Delta \theta$ etc.) and 
the total
available observing time, $t$. Secondly, a background
code keeps track of the current data on each event (from 
OGLE, MOA, RoboNet, and all teams that make data available in real time), 
and produces a new PSPL fit whenever new data
arrives. The results from these fits give the event parameters $\te$,
$A_0$ etc.~that are used to predict the magnification at
the requested time of observation. These two sets of inputs allow the
calculation of $g_i$ for each event using Eqn.~(\ref{eqn:goodness}), and
therefore an optimal list of targets with suggested exposure times
for the requested telescope at the requested time. The list is then
put out in either a machine readable or sortable human friendly
format. With RoboNet, this output controls the telescope, and new
data is fed back into the PSPL model to close the loop and give
priorities that are based on data just taken. For human observers at
other sites, the output pages are customisable to display any desired
parameters along with the priority of each microlensing event, and
also show light-curves and detection zone maps along with links to
the finding charts and original OGLE and/or MOA pages for each.
Although written for RoboNet, this prioritisation tool is freely
available and other microlensing observers are encouraged to make use
of it.

\subsection*{Acknowledgements}

Keith Horne was supported by a PPARC Senior Fellowship during
the early stages of this work.
We thank Steve Kane, Martin Dominik, Scott Gaudi, and Pascal Fouque
for helpful comments on early versions of the manuscript.

\bsp

\end{document}